\documentclass[a4paper,fleqn,usenatbib]{mnras}

\usepackage[T1]{fontenc}
\usepackage{ae,aecompl}

\usepackage{graphicx}	
\usepackage{amsmath}	
\usepackage{amssymb}	
\usepackage{tablefootnote}

\title[Short title, max. 45 characters]{A pebbles accretion model with chemistry and implications for the solar system}

\author[M. Ali-Dib]{
Mohamad Ali-Dib$^{1}$$^{,}$$^{2}$\thanks{E-mail: m.alidib@utoronto.ca}
\\
$^{1}$Centre for Planetary Sciences, Department of Physical \& Environmental Sciences, University of Toronto at Scarborough,\\
Toronto, ON M1C 1A4, Canada\\
$^{2}$Canadian Institute for Theoretical Astrophysics, 60 St. George St, University of Toronto, Toronto, ON M5S 3H8, Canada\\
}

\date{Accepted XXX. Received YYY; in original form 02 July 2016}

\pubyear{2016}

\begin{document}
\label{firstpage}
\pagerange{\pageref{firstpage}--\pageref{lastpage}}
\maketitle

\begin{abstract}
We investigate the chemical composition of {the solar system's} giant planets atmospheres using a physical formation model with chemistry. The model incorporate disk evolution, pebbles and gas accretion, type I and II migration, {simplified} disk photoevaporation and solar system chemical measurements. We track the chemical compositions of the formed giant planets and compare them to the {observed values}. Two {categories of} models are studied: with and without disk chemical enrichment via photoevaporation. Predictions for the Oxygen {and Nitrogen} abundances, core masses, and total amount of heavy elements for the planets are made {for each case}. We find that in the case without disk PE, {both Jupiter and Saturn will have a small residual core and comparable total amounts of heavy elements in the envelopes. We predict oxygen abundances enrichments in the same order as carbon, phosphorus and sulfur for both planets. Cometary Nitrogen abundances does not allow to easily reproduce Jupiter's nitrogen observations.} In the case with disk PE, less core erosion is needed to reproduce the chemical composition of the atmospheres, so both planets will end up with {possibly more massive residual cores, and higher total mass of heavy elements}. {It is also significantly easier to reproduce Jupiter's Nitrogen abundance.} No single was disk was found to form both Jupiter and Saturn with all their constraints in the case without photoevaporation. No model was able to fit the constraints on Uranus \& Neptune, hinting toward a {more complicated formation mechanism for these planets. The predictions of these models} should be compared to the upcoming Juno measurements to better understand the origins of the solar system giant planets.
\end{abstract}



\section{Introduction}
Twenty years ago, the Galileo probe dove into Jupiter's atmosphere bringing a wealth of informations on its properties, structure and chemical composition. One of Galileo's significant results was the discovery that Jupiter's atmosphere is enriched by a factor of 2-4 with respect to solar value with most volatile elements \citep{atreya2003}, including noble gases and Nitrogen, {elements that condense} at extremely low temperatures. This same trend is also seen in Saturn, Uranus and Neptune, suggesting further that the solar system giant planets all formed via {core accretion, since the alternative gravitational instability model predicts solar abundances for all elements. A {summary} of {the giant} planets chemical compositions {and core masses} can be found in table \ref{table_enri} (from \cite{mousis2014}).}
 
\renewcommand\arraystretch{1.2}
\begin{table}
\begin{center}
\caption{Measured chemical enrichments in the giant planets atmospheres normalized to the protosolar value, followed by the observational constraints on their cores masses and total amount of heavy elements.}
\small
{\begin{tabular}{lcccccc}
\hline
\noalign{\smallskip}
				& \multicolumn{2}{c}{Jupiter}				& \multicolumn{2}{c}{Saturn}	 & \multicolumn{2}{c}{U \& N}			\\
Species			& E			& $\Delta$E$^{\rm(a)}$		& E			& $\Delta$E$^{\rm(a)}$ & E	\\
\hline
C				& 4.40		& 1.14					& 9.90		& 1.05				&	20-60\\
N				& 4.18		& 2.08					& 2.5--3.8		& --		& ??				\\
O$^{\rm(b)}$		& 0.42		& 0.15					& ??			& ??		& ??				 \\		
P				& 3.34		& 0.36					& 11.54		& 1.35	& ??				\\		
S				& 2.94		& 0.70					& 12.88		& 1.52	& ??				\\		
He				& 0.72		& 0.04					& 0.71		& 0.14	& ??				\\		
Ne$^{\rm(c)}$		& 0.12		& --						&??			& ??		& ??				\\		
Ar				& 2.62		& 0.86					&??			& ??		& ??				\\		
Kr				& 2.23		& 0.61					& ??			& ??		& ??				\\		
Xe				& 2.18		& 0.61					& ??			& ??		& ??				\\		
\hline		
D/H 				& 1  &								&1 		&	& 1.9 &	 \\
\hline
M$_c$ (M$_\oplus$) &  < 10  & & < 20    & &  ?? \\
M$_Z$ (M$_\oplus$) &  < 42  & &  16 - 30   & &  > 85\%     \\
\hline

\end{tabular}}\\
$^{\rm(a)}$Error is defined as ($\Delta$E/E)$^2$ =  ($\Delta$X/X$_{\rm planet}$)$^2$ + ($\Delta$X/X$_{\rm Protosun}$)$^2$; $^{\rm(b)}$this is a lower limit; $^{\rm(c)}$this is an upper limit.
\label{table_enri}
\end{center}
\end{table}

Even though the details of Jupiter's formation remain elusive, several mechanisms --in the framework of core accretion-- has been proposed as the origin of the observed enrichments :

1- The giant planets formation in the cold outer disk where {the temperatures are low enough for the volatiles to condense or get trapped in clathrates-hydrates or amorphous ices \citep{owen1999,gautier2001,hersant2004,alibert2005,mousis20122}}. These models usually assume high trapping efficiency of volatiles in the ices.

2- The giant planets formation in a chemically evolved disk \citep{guillot2006,throop,monga2015}. In these scenarios, the disk evaporation along with dust settling \& solids drift and sublimation enrich the midplane in {metals, where the giant planets will later form and accrete the enriched gas.}

The {chemical} enrichment mechanism in the giant planets and specifically Jupiter is therefore still an open question. More detailed discussions on these models can be found in \cite{mousis2014}.\\

In this work, we investigate these proposed mechanisms in the light of new observational and theoretical advancements in planets formation and protoplanetary disks chemistry. For the first time, Argon and molecular Nitrogen abundances have been measured in a comet (67P/C-G by Rosetta \citep{rubin2015,balsiger2015}) giving us valuable insights on the composition of the protosolar nebula. Moreover, the in the last few years, the fast and efficient pebbles accretion models became main stream scenarios for planets formation \citep{lamb1,lamb2}, giving promising solutions to long standing problems on the origin of giant planets. This type of accretion for example can allow for a rapid formation of a giant planet's core even in the outer tenuous nebula. Finally, the \textit{Juno} space probe currently {around} Jupiter will will tentatively measure the elusive Oxygen abundance of the planet in addition to refining its nitrogen abundance and core mass values \citep{bolton2010}. It is crucial hence to have quantitative predictions for these quantities from the different enrichment scenarios (formation in the outer disk vs enrichment via photoevaporation) that are compatible with the known planets formation physics in addition to the chemical constraints we have on the giant planets. 

The questions {we address in this work are:}
\begin{itemize}
\item Where/when should Jupiter and Saturn form if their chemical compositions were acquired from a disk with solar composition ?
\item How much should the initial solid core get eroded into the envelopes of these planets to reach the observed chemical enrichments ?
\item How will the disk chemical enrichment via photoevaporation affect the above ?
\item Can Jupiter and Saturn still form late in the disk when photoevaporation has {started}?
\item Are these models compatible with the dynamical histories of these planets ?
\item Can we disentangle these formation scenarios via Juno measurements ? 
\end{itemize}

In section 2 we detail the planets formation model and our approach to disk chemistry. We discuss the results in section 3 and finally summarize and conclude in section 4. \\


\section{Pebbles accretion model with chemistry}
Multiple works have tried to couple the disk's chemistry to a planet formation model. \cite{dr1,dr2} for example coupled a classical core accretion model along with gas-grain chemistry to study the structure of the disk and the formation of Uranus and Neptune. \cite{alibert1,alibert2} on the other hand coupled the population synthesis model of \cite{alibert3} to an equilibrium chemistry module in order to to study the C/O and Mg/Si ratios in exoplanets. Both of these approaches are classical, they include core formation via planetesimals accretion and not the new paradigm of pebbles accretion. They also both use chemical models, while a plethora of observations is available for the solar system. The underlying idea of this work is hence to use solar system and disks chemistry observations along with a pebbles accretion in order to statistically constraint the disk and planetary initial conditions fitting the measurements. The models that fit best can then be used to make predictions for Juno.  

\subsection{Disk \& Formation}
\subsubsection{Gaseous disk}
Underlying our model are protoplanetary disks {structure profiles} fits for the elaborate simulations of \cite{bitsch2} (as detailed in their appendix A and provided as a script by B. Bitsch). The 2D (radial and vertical) accretion disk model of \cite{bitsch1,bitsch2} feature both viscous and direct stellar heating along with realistic radiative cooling and opacities treatment, generating bumps and dips in the structure of disk where planets can preferentially {grow due to their migration slowing down.} We refer the reader to \cite{bitsch1,bitsch2} for more detailed informations on these models. The free parameters of the fit script are the disk's metallicity ``\textit{metal}''\footnote{we will always follow pre-set notations from the cited papers for easiness of comparison.}, turbulence parameter ``$\alpha$'', and the disk mass accretion rate related to the T-tauri star age through the \cite{Hartmann1998} correlation:
\begin{equation}
log\bigg(\frac{\dot{M}}{M_{\odot}/yr}\bigg)=-8.0-1.40\times log\bigg(\frac{t}{10^6 yr}\bigg)
\end{equation}
we follow \cite{bitsch2,bitsch3} in fixing $\alpha$ to 0.0054 to remain consistent with their results and for the disk script to remain faithful to the full simulations. we also fix the initial \textit{metal} to the nominal 0.5\% following \cite{bitsch3}, but let it evolve with the disk photoevaporation when taken into account. The script will hence generate disk models as a function of time with constant metallicity and turbulence. This imply that all of the disk parameters in our simulations are fixed, leading to a {constrained free parameters space} limited to the planets initial conditions instead of the disk's.

\subsubsection{Initial seeds \& Pebbles disk}
In the core accretion model of planets formation, a giant planet core starts as a small ``seed'', then accrete more solids {till it grow massive enough} to accrete gas \citep{pollack1996}. The current paradigm for planetary seeds formation is via gravitational collapse assisted by streaming instability \citep{youdin2005,johansen2007}. These models found seeds with masses in the order of dwarf planet Ceres (10$^{-4}$ M$_\oplus$). We hence will explore the effect of different initial seeds masses {in the order of} this value {injected} in the disk. Once a seed has formed, it will start growing via accretion. The timescale of this accretion has always been problematic since in the classical models where cores accrete large kilometric planetesimals, growth timescales were consistently found to be longer than the disk lifetime. A promising solution proposed initially by \cite{lamb1} consist of growing the cores through small pebbles strongly coupled to the gas. These pebbles feel the gas drag and drift efficiently onto the cores. This has came to be known as pebbles accretion scenarios. \cite{lamb2,lamb3} pushed this model forward proposing that small dust in the outer disk, beyond the so called ``pebbles production line'', grow by coagulation into pebbles that drift inward till getting accreted by planetary cores, and showed how this leads to an inside out formation of giant planets with architecture resembling our own solar system. In this work we follow \cite{lamb3,bitsch3} in expressing the pebbles production line as:
\begin{equation}
r_g=\bigg(\frac{3}{16}\bigg)^{1/3}(GM_\star)^{1/3}(\epsilon_D Z)^{2/3}t^{2/3}
\end{equation}
and
\begin{equation}
\frac{dr_g}{dt}=\frac{2}{3} \bigg(\frac{3}{16}\bigg)^{1/3}(GM_\star)^{1/3}(\epsilon_D Z)^{2/3}t^{-1/3}
\end{equation}
where $Z$ is the disk metallicity in pebbles and $\epsilon_D=0.05$ is a parameter constrained from advanced coagulation simulations.
The pebbles flux is then defined as:
\begin{equation}
\dot{M}_{peb}=2\pi r_g \frac{dr_g}{dt}(Z\Sigma_g)
\end{equation} 
where $\Sigma_g$ is the gas surface density. Finally the pebbles surface density is defined as:
\begin{equation}
\Sigma_{peb}=\sqrt{\frac{2\dot{M}_{peb} \Sigma_g}{\sqrt{3} \pi \epsilon_P r_P v_K}}
\end{equation}
where $\epsilon_P=0.5$ and $v_K=\Omega_K r$.

\subsubsection{Pebbles accretion}
For {pebbles} accretion to start, planetary seeds massive enough need to be present in the disk. \cite{bitsch3} chose to start their simulations with seeds that already reached the pebbles transition mass, where pebbles accretion occurs in the Hill radius of the planetary seed. we choose however to explore a range of planetary seeds masses. At the lower end of this range, these seeds will grow through the inefficient Bondi accretion regime where pebbles are accreted only within the Bondi radius of the planet. These will grow till reaching the pebbles transition mass where the much more efficient accretion within the Hill radius resumes. we follow \cite{morby1} in modeling these processes.
The Bondi and Hill radii are defined as respectively:
\begin{equation}
R_B=\frac{GM}{(\Delta v)^2}
\end{equation} 
\begin{equation}
R_H=r\frac{M_c}{(3M_\odot)}
\end{equation}
where $\Delta v= \eta V_K$, with $\eta =- \frac{1}{2} \bigg(\frac{H}{r}\bigg)^2 \frac{d log P}{d log r}$, and $M_c$ the core's mass. 
We then define:
\begin{equation}
R_{GP}=min(R_B,R_H)
\end{equation}
and the effective radius for accretion onto the planetesimal {in the Hill regime:}
\begin{equation}
r_{eff}=(\tau/0.1)^{1/3} R_{GP}
\end{equation}
{and in the Bondi regime :}
\begin{equation}
r_{eff}= \bigg(\frac{4 \tau_f}{t_B}\bigg)^{1/2} \times R_{GP}
\end{equation}
{where $t_B$ is the Bondi timescale. Assuming optimally coupled particles for an efficient Bondi accretion ($t_B$=$\tau_f$), this equation reduces to:}
\begin{equation}
r_{eff}= 2 \times R_{GP}
\end{equation}
Finally:
\begin{equation}
\dot{M}_{2D}=\frac{2r_{eff}v_{rel}\dot{M_F}}{4\pi r \tau \Delta v}
\end{equation}
and:
\begin{equation}
\dot{M}_{3D}=\dot{M}_{2D}\Bigg(\frac{\pi r_{eff}}{2\sqrt{2\pi}H_{pb}}    \Bigg)
\end{equation}
where $v_{rel}=max(\Delta v,v_{shear})$, and the pebbles scale height $H_{pb}=H_g \sqrt{\alpha/\tau}$

This pebbles accretion will then continue till reaching the pebbles isolation mass \citep{bitsch3}:
\begin{equation}
M_{iso}\approx 20 \bigg(\frac{H/r}{0.05}\bigg)^3 M_{Earth}
\end{equation}
A core beyond this mass will change the local gas pressure gradient, halting the pebbles flux and hence stopping accretion.

\subsubsection{Gas accretion}
Once pebbles accretion onto the core has stopped, the planet's envelope cool down allowing gas accretion to start. In the core accretion model, gas accretion occurs over two phases: the initial slow phase {when the envelope is less massive that the core, followed by the very rapid hydrodynamical collapse phase once the envelope had reached the core's mass.} we follow \cite{bitsch3} in modeling the {initial} slow accretion rate (based on the cooling timescales found by \cite{piso}) by:
\begin{equation}
\begin{aligned}
\dot{M}=0.00175f^{-2} \bigg(\frac{\kappa_{env}}{1 cm^2/g}  \bigg)^{-1}\bigg(\frac{\rho_{c}}{5.5 g/cm^3}  \bigg)^{-1/6}\bigg(\frac{M_{c}}{M_E}  \bigg)^{11/3} \\
\times \bigg(\frac{M_{env}}{0.1 M_E}  \bigg)^{-1} \bigg(\frac{T}{81 K}  \bigg)^{-0.5} \frac{M_E}{Myr}
\end{aligned}
\end{equation}
where $\kappa_{env}$ is the envelope's opacity regulating its cooling rate and thus the contraction speed, $M_{env}$ its mass, and $\rho_{c}$ and $M_c$ the core's density and mass. {The values we used for $\kappa_{env}$ and $\rho_{c}$ are shown in \ref{t2}}. {The gas contraction timescales in this model is a few 10$^5$ yr, slower than the $\sim$10$^6$ yr used in the classical core accretion models \citep{pollack1996}. This is due to the slightly reduced opacity value used, caused by the dust-pebbles dichotomy of the solids in the disk. }

To model the fast accretion phase, we also follow \cite{bitsch3} (based on: \cite{machida}), where they fit 3D hydrodynamical simulations as: 
\begin{equation}
\dot{M}_{gas,low} = 0.83\Omega_K \Sigma_g H^2 \bigg(\frac{r_H}{H} \bigg)^{9/2}
\end{equation}
and :
\begin{equation}
\dot{M}_{gas,high} = 0.14\Omega_K \Sigma_g H^2 
\end{equation}
where the low branch is for low mass planets with $\bigg(\frac{r_H}{H} \bigg) < 0.3 $.

\subsubsection{Disk migration}
To model disk migration we again remain consistent with \cite{bitsch3}. For low mass planets, Type I is implemented following \cite{paard1,paard2} taking into account both the Lindblad and corotation torques of the gas on the planet. The total torque is hence:
\begin{equation}
\Gamma_{tot} = \Gamma_L + \Gamma_C
\end{equation}
where both of them depend on the local density, temperature and entropy gradients in the disk. {The corotation torque's saturation is also included in this model.}

When a planet is massive enough to open a gap in the disk, {it becomes coupled to its viscous evolution and undergoes Type II migration.} The criteria for gap opening \citep{crida} is:
\begin{equation}
P=\frac{3}{4}\frac{H}{r_H}+\frac{50}{qR} \leqslant 1
\end{equation}
and the migration timescale is given by:
\begin{equation}
\tau_{II} = \tau_{\nu} \times max \Bigg(1, \frac{M_P}{4 \pi \Sigma_g r^2_P}      \Bigg)
\end{equation}

Type I migration can be slowed down by the opening of a partial gap before the planet is massive enough to undergo type II migration. We hence follow \cite{crida2007} and multiply the type I migration rate by :
\begin{equation}
f(p)=  
    \begin{cases}
      \frac{P-0.541}{4}, & \text{if}\ P<2.4646 \\
      1.0-exp(-\frac{P^{3/4}}{3}), & \text{otherwise}
    \end{cases}
\end{equation}

\subsubsection{Photoevaporation}
Photoevaporation have substantial effects on the physical and chemical evolution of disks. It is thought to be the leading mechanism dispersing disks on average 3 Myr after their formation. It is also thought to affect the disk's chemistry \citep{guillot2006,monga2015}.
The disk can be hydrodynamically photoevaporated through internal radiation via either the central star's EUV (the ``UV switch'') \citep{clarke2001} or via the much more powerful X-rays heating \citep{owen2010}. Both of these mechanisms lead to an inward out dispersal of the disks. In young disks, the surface is heated by radiation, but the evaporation rate is lower than inward accretion rate. Later, when accretion slows down, photoevaporation will dominate and start quenching the accretional flow beyond a certain radius where the gas is loosely bound. This will carve an inner hole in the disk, followed by the complete gas dispersal at a later time. In these models, solids in the evaporating layers of the disk can decouple from the gas and settle toward the midplane, enriching it in heavy elements. A disk can also be photoevaporated through the FUV radiation of nearby massive stars. This mechanism is outside in, where the disk outer edge will evaporate first, followed by the inner disk \citep{adams2004}. This mechanism seems to be favored for our own protosolar solar nebula due mainly to the unexplained edge of the Kuiper belt around 50 AU \citep{allen2001} (and other evidences reviewed in \cite{monga2015}). In the case of external photoevaporation, it is mostly solids drifting inward from the evaporating outer parts of the disk that will chemically enrich the planets forming regions. A detailed review of disks dispersal mechanisms can be found in \cite{alexander2014}. Properly simulating disks photoevaporation implies coupling photoevaporative winds to disk evolutions models, {which is} by itself numerically challenging. Adding the disk's chemical evolution to the problem adds multiple more layers of complexity. Coupling all of these to a planets formation and evolution model is numerically prohibitive. For these reasons we choose a simplified photoevaporation model to capture the essence of its physical and chemical effects. Modeling disk internal photoevaporation (UV switch and Xrays) necessitate intrinsically a full dynamical model due to the power balance between the evaporative flow, inward accretion and gas diffusion. This is beyond the scope of this work since the disk model we are using is simply mathematical fits to the original 2D model (which, again, does not simulate the viscous evolution of the disk). Modeling the external photoevaporation via the cluster's FUV radiation is easier to fit within the constructed model. In this case, since the disk is cleared outside in (in the same radial direction as accretion), we can crudely model this effect by simply reducing the photoevaporation mass loss rate $\dot{M}_{PE}$ from the initial mass accretion rate of the disk $\dot{M}_{acc}$. For $\dot{M}_{acc} \gg \dot{M}_{PE}$, the disk model is unaltered and reproduce the results of \cite{bitsch2}. For $\dot{M}_{acc} \sim \dot{M}_{PE}$, the photoevaporative flow takes over. we hence replace $\dot{M}$ of \cite{bitsch2} by:
\begin{equation}
\dot{M}=\dot{M}-\dot{M}_{PE}
\end{equation}
and then define the gas chemical enrichement as:
\begin{equation}
    \varepsilon_c= 1 + \frac{\dot{M}_{PE}}{\dot{M}_{acc}}
\end{equation}

we are hence implicitly assuming that 100\% of the heavy elements in the photoevaporated gas are retained. This entails that none of the solids are lost with the gas, and thus are large enough to feel the gas enough. Our model should hence be interpreted as an upper limit on the gas enrichment via photoevaporation. Moreover, this implies that volatiles elements are also preserved. This can happen through the mechanism proposed by \cite{monga2015} {the photoevaporating photons, with the help of mixing processes in the disk, can continuously generate enough water vapor in the outer cold nebula to trap most of the gaseous volatiles in icy particles.} For self consistency, we also multiply the disk's dust metallicity ($metal$) by $\varepsilon_c$, to account for the increase of the metallicity during photoevaporation. We limit $metal$ to $\leq$ 3 for the fits to remain accurate to the full disk model. Photoevaporation should also affect the metallicity in pebbles ($Z_0$), however this is not included in this model because quantifying this effect goes beyond the scope of this work. 

\subsection{Core mass \& erosion}
Measuring the core mass of giant planets is very challenging. For Jupiter, the only constraints we currently have are $M_c$ < 10 $M_{\oplus}$ and a total mass of heavy elements $M_Z$ between 10 and 42 $M_{\oplus}$ (\cite{guillot2014}, {and cf. Figure 8 of \cite{milit})}. In the case of Saturn, precise measurements using Cassini along with a {less dense} envelope allows a better estimate for $M_Z$, that was found to be between 16 and 30 $M_{\oplus}$. Its core mass however is even less constrained than for Jupiter, {since a high inner envelope metallicity is gravitationally hard to distinguish from a solid distinct core, so only an upper limit of 20 $M_{\oplus}$ can be inferred \citep{guillot2014}. In this work, any solids not well mixed in the envelope are supposed to be part of the solid core.}\\
The final chemical composition of a giant planet's atmosphere depends on the heavy elements abundance in the gas accreted during the final stage of core accretion, but equally on the fraction of the planet's core that got eroded and then mixed via convection in the envelope. Core erosion takes place when the temperature of the core-envelope boundary is higher than its thermodynamical stability limit. Evidences for core erosion in Jupiter and Saturn include, in addition to the heavy elements enrichment of their envelopes, their possibly low cores masses incapable of accreting a gaseous envelope in core accretion models.
{Modeling the core erosion require a detailed understanding of the interior structure of the planets and the heat \& matter transport mechanisms at play, which is beyond the scope of this work.} We will therefore take a simple approach parametrizing the erosion by introducing an erosion factor $0\leqslant f_E \leqslant 1$. The core is completely eroded for $f_E=1$, and is intact in the opposite case. When the core is partially or completely eroded, the removed fraction is added to the envelope, {thus increasing its metallicity}. The resulting core masses and atmospheric metallicity can then be compared to observations. In the simulations, we used a step size of 0.2 for this parameter for Jupiter and Saturn, and 0.05 for Uranus \& Neptune. {Throughout this manuscript, by ``core mass'' ($M_c$) we mean the mass of the solid core remaining and measurable today, after any initial core erosion.}   

\subsection{Planetary dynamics}
Another constraint on the solar system's giant planets is their dynamical configuration. In the Nice model \citep{n1,n2,n3,n4}, the four giant planets are thought to start close to one of their first order mean motion resonances after the disk dissipation, crossing this resonance later and triggering the global instability that shaped our solar system. One of the leading dynamical evolution models giving the suitable initial conditions for the Nice model from disk physics is the Grand tack scenario \citep{walsh2011}, where Jupiter undergoes Type II migration into the telluric planets region of the disk while a still forming Saturn goes through the much quicker (runaway) type III migration till getting locked with Jupiter in their mutual 2:3 MMR. It has been showed by \cite{masset2001} that this configuration will reverse the type II migration and lead to both planets migrating outward in a compact configuration. Moreover, \cite{pierens2008,pierens2011,pierens2014} showed using hydrodynamical simulations that this 2:3 resonance trapping followed by the outward migration is the most common outcome for the evolution of a Jupiter-Saturn like pair of planets. For all these reasons, we can use the crossing of the 2:3 resonance by a pair of Jupiter-Saturn mass planets as possible additional constraint on the solar system evolution. This is implemented in the model as a simple yes/no indicator showing if this resonance was crossed by Saturn while its mass is higher than 50 $M_{\oplus}$ \citep{pierens2008}. It should be noted though that type III migration is not implemented in this model, so this should be seen as a check if type I and II migration can lead the planets crossing this resonance with the right chemical composition.

\subsection{Chemistry}

\subsubsection{The giant planets region chemical composition}
The final chemical composition of a planet will depend on the amount of metals it accreted during its formation. If a chemical element is present in the nebula purely in gaseous phase, its abundance with respect to the similarly gaseous hydrogen will remain constant. The element will therefore have a solar abundance. A chemical specie hence needs to be present at least partially as a solid in the nebula to get enriched with respect to the solar value. A chemical element can either be in solid phase at all relevant temperatures (refractory elements) or condense at low enough temperatures (volatiles). The major volatile in protoplanetary disks, water, condense around 150 K. The other important volatiles found in the disk have substantially lower condensation temperatures and hence farther snowlines \citep{fray}. Two types of bodies with substantial amounts of the relevant chemical species are thought to have formed in the giant planets region: C-type asteroids and Oort cloud comets (OCC). C-type asteroids are thought to be the parent bodies of carbonaceous chondrites, and thus the composition of these meteorites is used as a proxy to understand {the physico-chemical structure of this region}. To date, 1P/Halley is the only OCC studied in situ, and thus is our only proxy to the refractory composition of this cometary family. 67P/C-G and 81P/Wild have also been studied respectively in situ by Rosetta and via sample return, but both are Jupiter Family Comets (JFC) that probably originated in the Kuiper Belt. The following is a summary of the current knowledge on the chemistry of the giant planets formation region.
\subsubsection{Oxygen}
Oxygen is the most abundant metal in most disks. In our own protosolar nebula with a C/O ratio of 0.55, oxygen is twice as abundant as carbon, and thus the disk environment was on average oxidizing. Equilibrium chemistry calculations \citep{cyr1999} show that the oxygen will be distributed between refractories (Mg and Si bearing elements) and volatiles (mainly water). In situ observations in comet Halley \citep{jessberger1988} showed that 43\% of the oxygen is in the refractory phase, while the remaining 57 \% are in {water}. These are the values we will use in this models. 
\subsubsection{Carbon}
On one hand, comets observations show that C bearing volatiles (CO and CO$_2$) can only account for up to 50\% in molar abundance with respect to the solar value, although the value varies from a comet to another and can be much lower \citep{mumma}. This implies That the remaining carbon (at least $\sim$ 50\% of solar molar abundance) must be in the cometary carbonaceous refractories. In Halley, when normalized to Mg, 80\% of carbon was found to be in the refractory dust \citep{jessberger1988, encrenaz}. On the other hand, CI chondrites are depleted in carbon (in refractory phase) on average by an order of magnitude with respect to the solar value \citep{lodders2003}. It is however unlikely that the giant planets have accreted significant amounts of CI chondrites-like materials, because this contradicts their carbon rich nature. In the sample returned from comet 81P/Wild, no CHON particles were found, but this is attributed to a measurement error caused by capture heating \citep{zolensky2006,2011IAUS..280..261B}.
For these reasons, we will follow {Halley} measurements in assuming that 80\% of the solar carbon abundance in the disk is in refractories, and the rest in the volatile CO. The volatile component will be in gaseous phase inside the CO iceline at 22 K and in ices beyond it. {It should be mentioned that since we are also fitting the nitrogen and argon abundance in both Jupiter and Saturn, the amount of carbons in refractories will not significantly affect the the results. This is because the N$_2$ condenses at slightly lower temperatures than CO. Therefore, our results are more sensitive to the amount of nitrogen in refractories.}
\subsubsection{Nitrogen}
The nature of the main nitrogen reservoir in the solar system bodies is controversial. In CI chondrites, nitrogen (in its refractory phases) is depleted by 2 orders of magnitude with respect to the solar value. One would then expect the volatiles N$_2$ and NH$_3$ to be the major nitrogen bearing species in the nebula. However, NH$_3$ abundance in comets is only a fraction of the solar value \citep{mumma}, and N$_2$ was recently found to be very depleted {by more than an order of magnitude} in comets \citep{rubin2015}, even though it has the same volatility as the much more abundant CO \citep{fray}. For these reasons, we will assume that nitrogen in the nebula is present mostly as the super volatile N$_2$, with only 14 \% of its molar abundance in refractory phase (the value found in Halley's dust \citep{jessberger1988}). we assume further an N$_2$ condensation temperature of 20 K \citep{fray}. 
\subsubsection{Sulfur}
The abundance of sulfur in the solar system small bodies varies wildly. Refractory sulfur in CI chondrites has an almost solar abundance \citep{lodders2003}, implying a low fraction in volatiles. In comets, the volatile phase of sulfur (H$_2$S) represent at most up to 2/3 the solar abundance, although (as in the case of volatile carbons) can be much lower. In Halley though, all sulfur was found to be in refractories reaching up to almost 1.5$\times$ solar abundance. This anomalous value is usually interpreted as measurement error due to ion yielding issues in the data analysis, making these results less reliable. Sulfur was also detected in the dust samples returned from Wild 2 by Stardust and found to be $\sim$ 0.17 times the CI chondrites value \citep{flynn2006}, much lower than in Halley and too low to account for all of the solar sulfur abundance along with volatiles. Wherever this value is real or again a measurement error is not settled. With all this uncertainty, we are going to assume that all of the disk's sulfur is in refractory form, neglecting the sulfur found in volatiles. This assumption should have a minimal impact on the results since the volatile H$_2$S condensate at the relatively high temperature around 100 K. 

\subsubsection{Phosphorus}
Phosphine (PH$_3$, the main volatile phase of phosphorus), is not yet detected in comets, with only upper limits available (\cite{phosphinepaper} and references herein). It was though detected in a circumstellar region, and found to contain only 2\% of the solar phosphorus abundance \citep{phosphine2}. On the other hand, the refractory phases of phosphorus in CI chondrites have an almost solar abundance\citep{lodders2003}. We will hence assume that all of the phosphorus in the disk is in refractory phase.

\subsubsection{Argon}
Argon, an ultra-volatile noble gas, is hard to find in the solar system. In chondrites it is depleted by multiple orders of magnitudes with respect to the solar value. This is expected since all chondrites (mainly refractories) are very poor in the volatile phases of chemical species. Argon's abundance in ices is less clear however. Recently, first detection of Argon in a comet (67P/C-G by Rosetta \citep{balsiger2015}) translated into an relative abundance with respect to water of 0.1 - 2.3 $\times 10^{-5}$, again orders of magnitude below solar value. This questions the formation of Jupiter from mainly 67P-like material, since this would have lead to an Argon depleted Jupiter which is not the case. Another possibility is the chemical evolution of the comet since its formation, but a caveat to this idea is the much higher abundance of the equally volatile CO. For all these reasons, we assume that all of the Argon in our disk is in free gaseous form. The Argon iceline is assumed to be around 24 K.

\subsubsection{Other elements}
Since most other volatile elements (H, He, Ne, Kr and Xe) are extremely depleted (or never detected) with respect to solar value in both comets and CI chondrites, we assume that they are absent from the solid phase and present only in gaseous form inside their snowlines with a solar abundance throughout the disk.

\subsubsection{D/H ratio}
The Deuterium/Hydrogen (D/H) ratio of a giant planet's atmosphere is affected by two factors: the D/H ratio of the accreted nebular gas, which should have a solar value, and the amount of highly deuterated ices dissolved into the envelope through core erosion. We hence {follow \cite{lecluse,feucht99}} in calculating a planet's D/H ratio as:
\begin{equation}
D/H_{p}=\frac{(1-x_{H_2})\times D/H_{ices}+x_{H_2}\times D/H_{H_2}}{D/H_{H_2}}
\end{equation}
where $D/H_{H_2}=2.25 \times 10^{-5}$ \citep{lellouch}, $D/H_{ices} = 20 \times D/H_{H_2}$ {(average cometary value from \citep{altwegg})} and $x_{H_2}$ is the volumetric ratio of $H_2$ defined as:

\begin{equation}
x_{H_2}=\frac{1}{1+\frac{1-f_{H_2}}{(m_{H2O}/m_{H_2})\times f_{H_2}}}
\end{equation}
with $f_{H_2}$ the mass ratio of ${H_2}$ and ${H_2O}$:
\begin{equation}
f_{H_2}=\frac{0.747 \times M_{H_2+He}}{0.747 \times M_{H_2+He} + M_{ice}}
\end{equation}

$D/H_{p}$ will be hence tend to the value $D/H_{H_2}$ for low core erosion (or Hydrogen dominated atmosphere such as Jupiter or Saturn), and tend to the value of $D/H_{ices}$ for high core erosion and low envelope masses (Neptunian planets).

\subsubsection{Calculating the chemical enrichment}
In this model the chemical enrichment of each element is tracked separately over each time step. we define $X$ as the enrichment over the solar value of a chemical element. $X$ is then $\geqslant 1$, where unity implies solar abundance. Over each time step, the planet core mass will increase by $dM_c$, thus increasing the enrichment by: 
\begin{equation}
dX  = \frac{dM_c}{\sum\limits_i^{} M_i n_i^\odot}\times (f_I+f_R)n_X^{\odot}
\end{equation}
where $M_i$ and $n_i^\odot$ are respectively the molar mass and relative solar molar abundance (with respect to $H_2$) of the element $i$, and $f_I$ \& $f_R$ are respectively the ices and refractories fractions of the chemical element. We can finally write:
\begin{equation}
X=1+ E_f\int\limits_{t_0}^{t_f} dX dt
\end{equation}

\section{Results}
we setup the simulations as a population synthesis in order to explore the parameters space. However, since there is a large number of tunable parameters in this model, but not all of them are expected to directly affect the planets chemistry, we choose to only vary the relevant parameters. All of the model's tunable parameters are shown in Table \ref{t2} along with their ranges and step sizes. The simulations are stopped when either time reaches 3 Myr, the disk photoevaporates completely, {or a planet the mass of Jupiter/Saturn/Neptune is reached.} {We summarize the relationship between the different observational constraints we have on the giant planets and the model's free adjustable parameters (varied in the population synthesis) in table \ref{summary}.} 
\subsection{No photoevaporation}
To understand the effects of the individual model parameters lets first first examine the case without photoevaporation, where the chemistry is completely determined by the core erosion.

\renewcommand\arraystretch{1.2}
\begin{table*}
\begin{center}
\caption{The range of values for the parameters used in the population synthesis. When no range was precised, a constant value was used for all simulations.}
\scriptsize
{\begin{tabular}{lcccc}
\hline
\noalign{\smallskip}

Parameters			& Range			& Step	& Range			& Step	\\
\hline
& \multicolumn{2}{|c|}{Single planet} & \multicolumn{2}{|c|}{Multiple planets}  \\
\hline
T$_{ini}$			& 10$^5$ - 2$\times$10$^6$ yr		& 10$^5$	 yr	 &   10$^5$ - 2.3$\times$10$^6$   &	2$\times$10$^5$ yr	\\
R$_0$	& 1.25 - 35 AU		& 1.25 AU	& 2.5 - 35 AU		& 2.5 AU				 \\		
M$_0$				& 1 $\times$ 10$^{-4}$ - 9$\times$ 10$^{-4}$ M$_\oplus$		& 2$\times$ 10$^{-4}$	& 1$\times$ 10$^{-4}$ - 9$\times$ 10$^{-4}$ M$_\oplus$		& 2$\times$ 10$^{-4}$			\\		
E$_f$				& 0 - 100\%  	& 20\%		& 0 - 100\%  	& 20\%		\\		
\hline
$\dot{M}_{FUV}$ (M$_{\odot}$/yr) &$5 \times 10^{-10}$ \&  $1 \times 10^{-9}$ -  $3 \times 10^{-9}$ & $1 \times 10^{-9}$ \\
metal & 0.5 \% & - \\
Z$_0$ & 1 \% & -  \\
$f$ & 0.2 & - \\
$\kappa_{env}$ & 0.05 cm$^2$/g & - \\
$\rho_c$ & 5.5 g/cm$^3$ & - \\
\hline		
C$_{ref}$ & 80 \% & -  \\
O$_{ref}$ & 43 \% & -  \\
N$_{ref}$ & 14 \% & -  \\
S$_{ref}$ & 100 \% & -  \\
P$_{ref}$ & 100 \% & -  \\
D/H$_{ice}$ & 20 $\times \ 2.25 \ \times 10^{-5}$  & - \\
\hline
H$_2$O {iceline} & 150 K & -  \\
Ar {iceline} & 24 K & -  \\
CO {iceline} & 22 K & -  \\
N$_2$  {iceline} & 20 K & -  \\
\hline

\end{tabular}}\\

\label{t2}
\end{center}
\end{table*}

\subsubsection{Single planet: Jupiter}
\begin{figure}
	\includegraphics[scale=0.2]{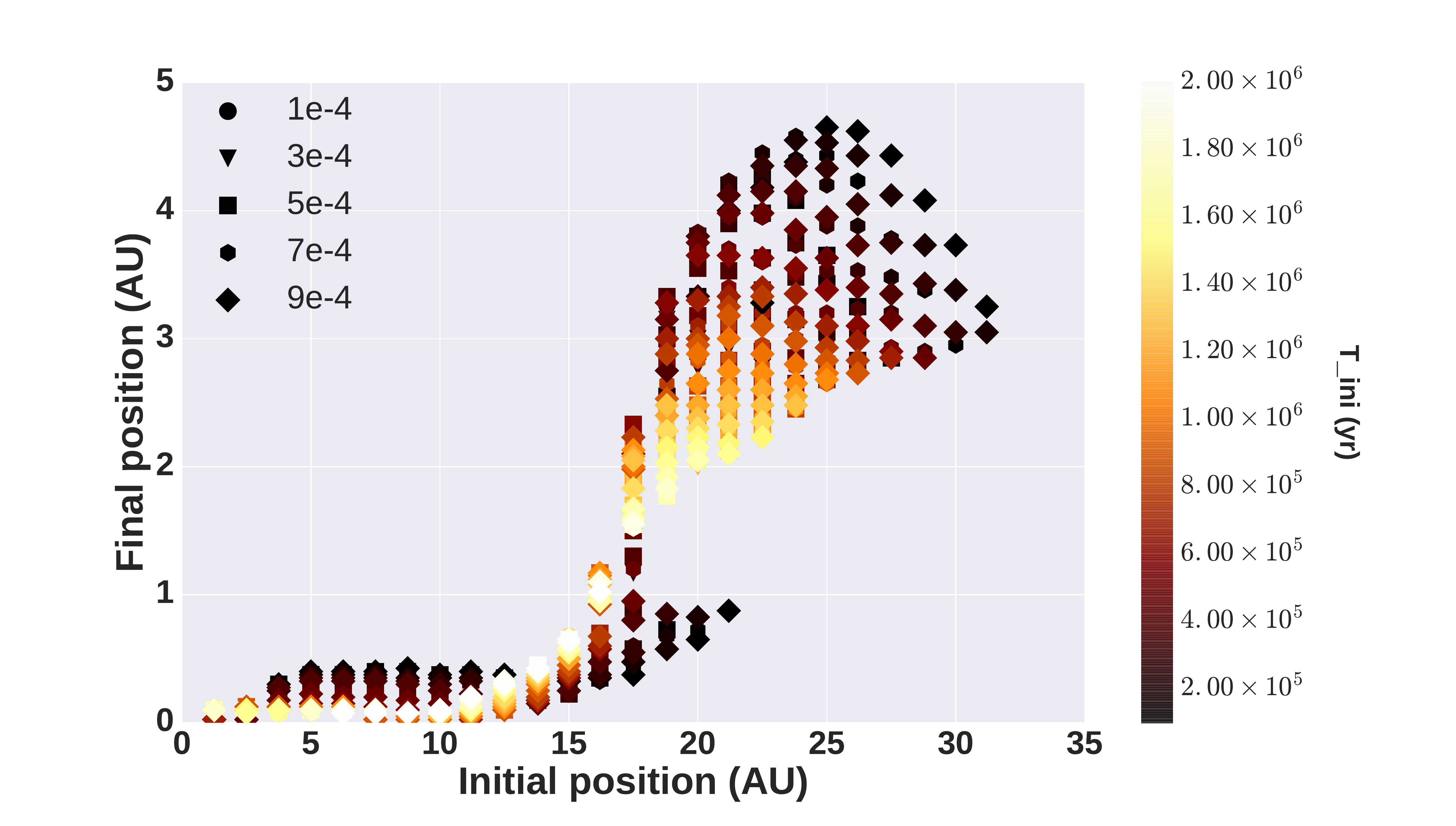}
   \caption{The final position (Rf) of Jupiter mass planets as a function of the seed's initial position (R0) and injection time (T$_{ini}$). The geometrical forms are the seed's initial mass in M$_{\oplus}$. This is the case with no disk photoevaporation, and showing all planets with no constraints.}
    \label{fig:jup1}
\end{figure}

\begin{figure}
	\includegraphics[scale=0.2]{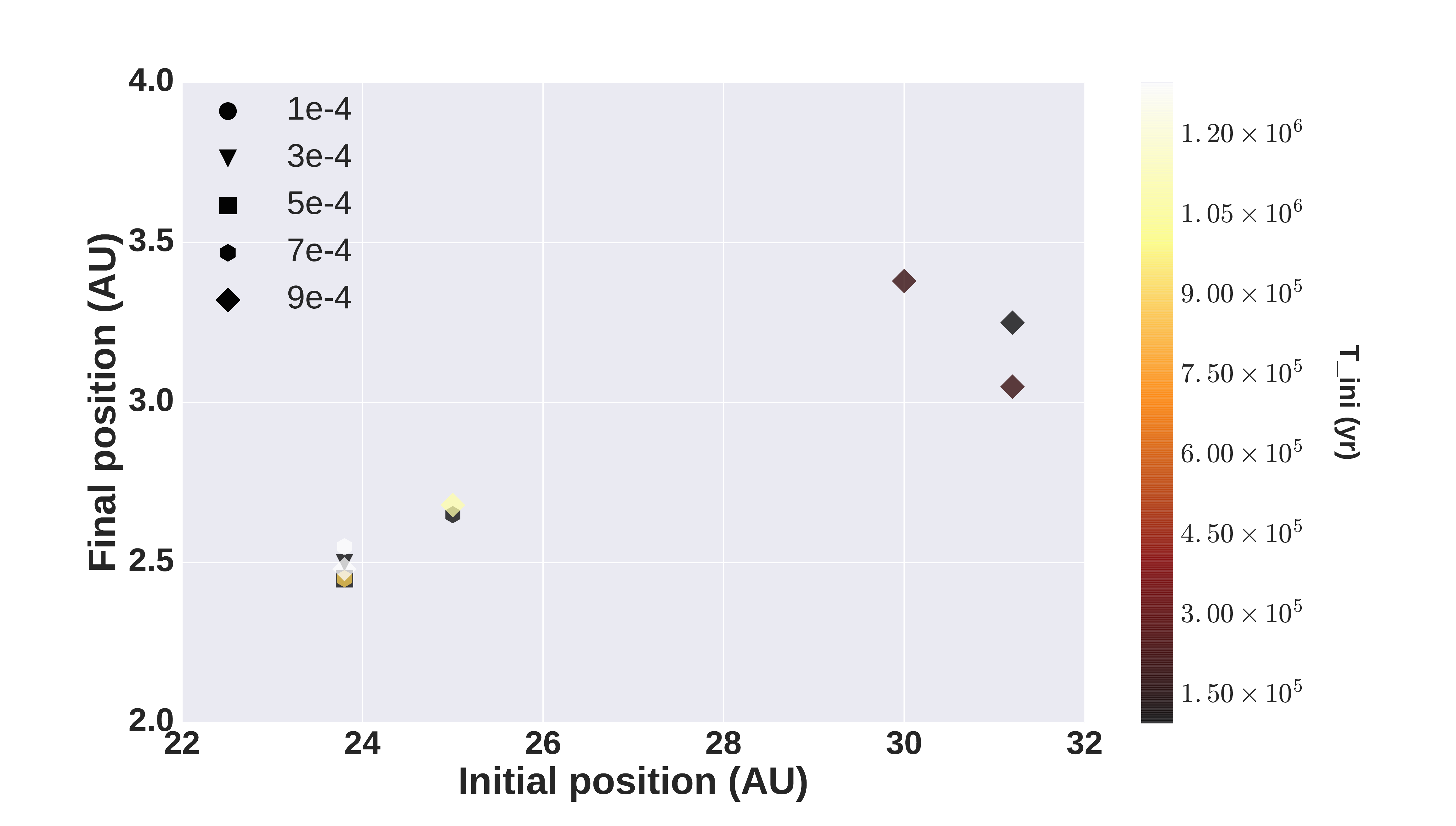}
   \caption{Same as Figure \ref{fig:jup1}, but with all constraints (chemical composition and core mass) applied.}
    \label{fig:jup2}
\end{figure}

\begin{figure*}
	\includegraphics[scale=0.2]{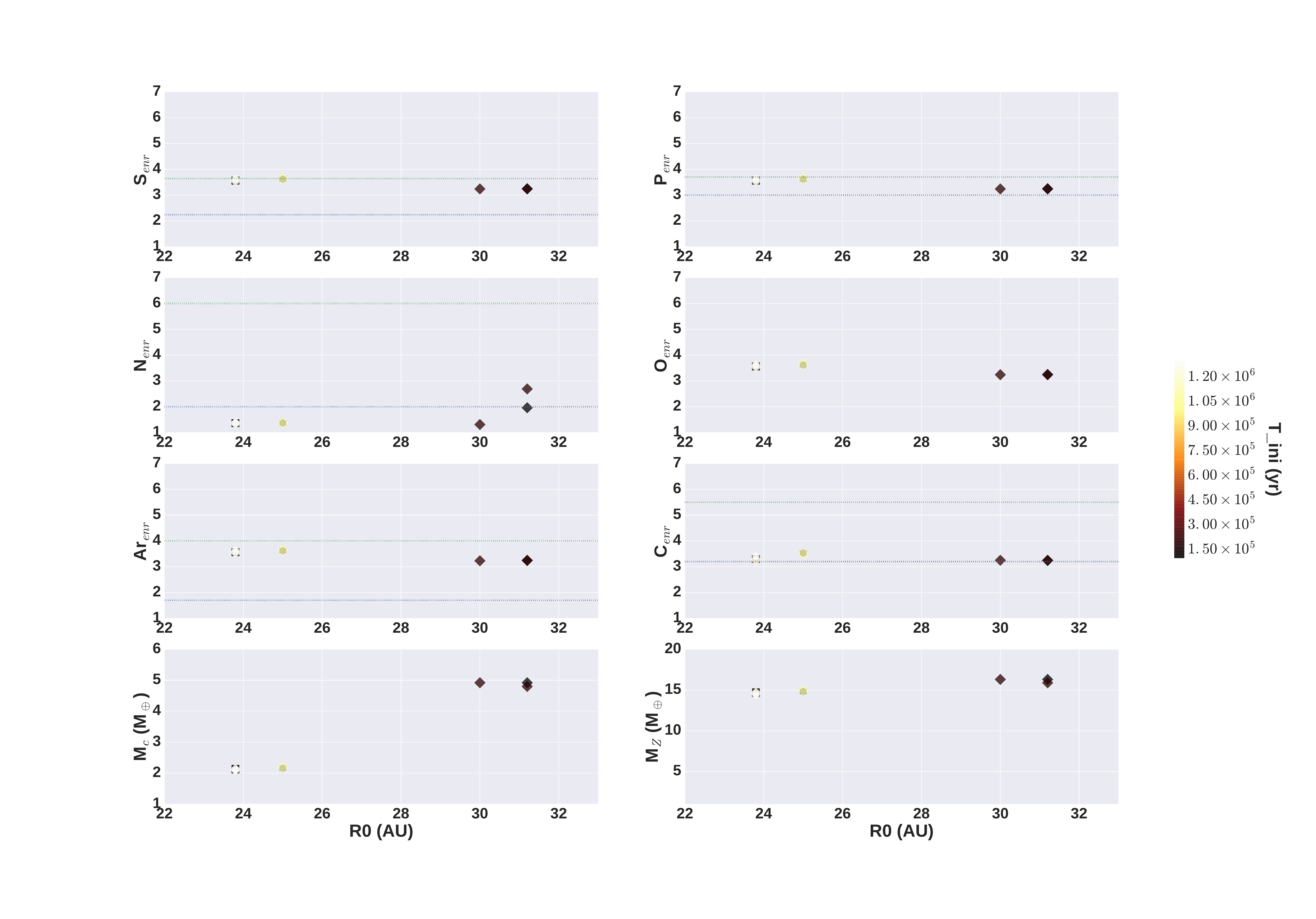}
   \caption{The enrichment factor of Sulfur, Phosphorus, Nitrogen, Oxygen, Argon and Carbon predicted by this model for the case with no disk photoevaporation for Jupiter. Bottom panels show the predicted core mass and total amount of heavy elements in the planet. The dashed horizontal lines are the observational constraints. }
    \label{fig:jup3}
\end{figure*}

Figure \ref{fig:jup1} shows the Jupiter population found with no chemical or core constraints. This plot shows, for every planet with Jovian mass, its final position as a function of the seed's initial position, mass, and time of injection in the disk. More than 10$^4$ planets where found to fit this one simple criteria. The one conclusion to take from this Fig. \ref{fig:jup1} is that even with detailed disk and migration modeling, only planets that started forming beyond 20 AU can end up in the 2-3 AU region, while the rest will become warm/hot Jupiters, as found by \cite{bitsch3}. Applying Jupiter's core mass constrain does not help a lot, reducing the number of matches by 10\%. Things gets more interesting when we add the abundances of Sulfur and Phosphorus. These elements are present throughout the disk only as refractories, and their abundances in Jupiter are known with a relatively low uncertainty. This implies that, given a certain initial core mass, these 2 elements can be used to constrain the core erosion factor of a giant planet. Applying these constrains decrease the number of matches {by 85\%}, while keeping the plot shape similar to Fig. \ref{fig:jup1}, simply because these indicators are not sensitive to the formation location of the planet. A large portion of the core mass/core erosion parameter space has been eliminated through this process. We now apply the remaining (location sensitive) constraints: {C and Ar}. This will leave us with only a handful of matches. {When we apply the nitrogen constraint though, only a single matching planet remain. For this reason, we will ignore the nitrogen abundance for the non photoevaporation case. The cause for this discrepancy can be either the low fraction of nitrogen in refractories we used, implying that during the planets formation phase the nitrogen refractory fraction was higher than what we find today in comets and chondrites, or that nitrogen condensed at a higher temperature than the value we used for the N$_2$ iceline, possibly because it was trapped in clathrates.} The final matching planets positions and initial conditions are shown in Fig. \ref{fig:jup2}. The detailed properties of the planets are shown in Fig. \ref{fig:jup3}. All these planets, by definition, fits all of the chemical constrains shown in the plot. They all formed between 22 and 32 AU, to allow the accretion of enough Ar and CO to match the measurements, and they all have core erosion factors of {60-80\%}, meaning that only {20-40\%} of their initial solid core remains intact. {Another characteristic shared by all but two of these planets is that the initial seed mass was $7-9 \times 10^{-4} M_{\oplus}$, the highest values used, {implying that a seed significantly more massive than Ceres is needed to form a giant planet}. The remaining 2 cases had initial seed masses of 3 and 5 $\times 10^{-4} M_{\oplus}$. This trend also holds for Saturn. {Finally, all of the final matching planets started forming relatively early in the solar nebula (no later than 1.2$\times$10$^6$ yr). This is simply because of the constraints on the core mass via the phosphorus and sulfur proxies. Since the core's isolation mass scales with the aspect ratio of the disk, and this quantity decreases with time, only planets that formed early in the disk can acquire the necessary amounts of heavy elements to match observations. A planetary core should also have enough time to grow sufficiently large via the slow Bondi regime to start accreting pebbles.  \\}
This model makes precise predictions, shown in table \ref{t3} for the abundance of oxygen in Jupiter (3-4 $\times$ solar), in addition to its current core mass (2 or 5 $ M_{\oplus}$) and total mass of heavy elements (15-17 $ M_{\oplus}$).}

\subsubsection{Single planet: Saturn}
\begin{figure}
	\includegraphics[scale=0.2]{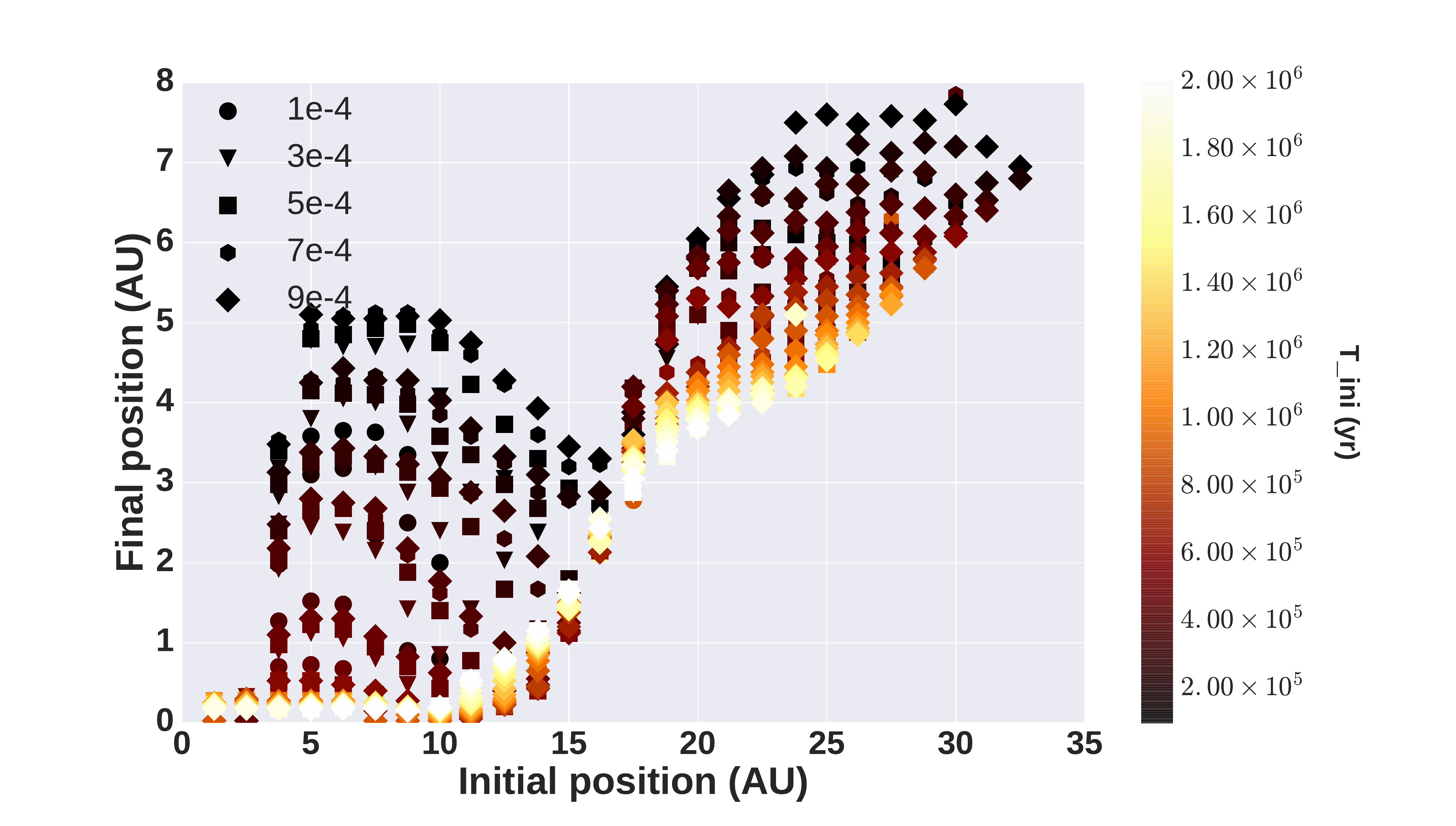}
   \caption{The final position (Rf) of Saturn mass planets as a function of the seed's initial position (R0) and injection time (T$_{ini}$). The geometrical forms are the seed's initial mass in M$_{\oplus}$. This is the case with no disk photoevaporation, and showing all planets with no constraints.}
    \label{fig:sat1}
\end{figure}

\begin{figure}
	\includegraphics[scale=0.2]{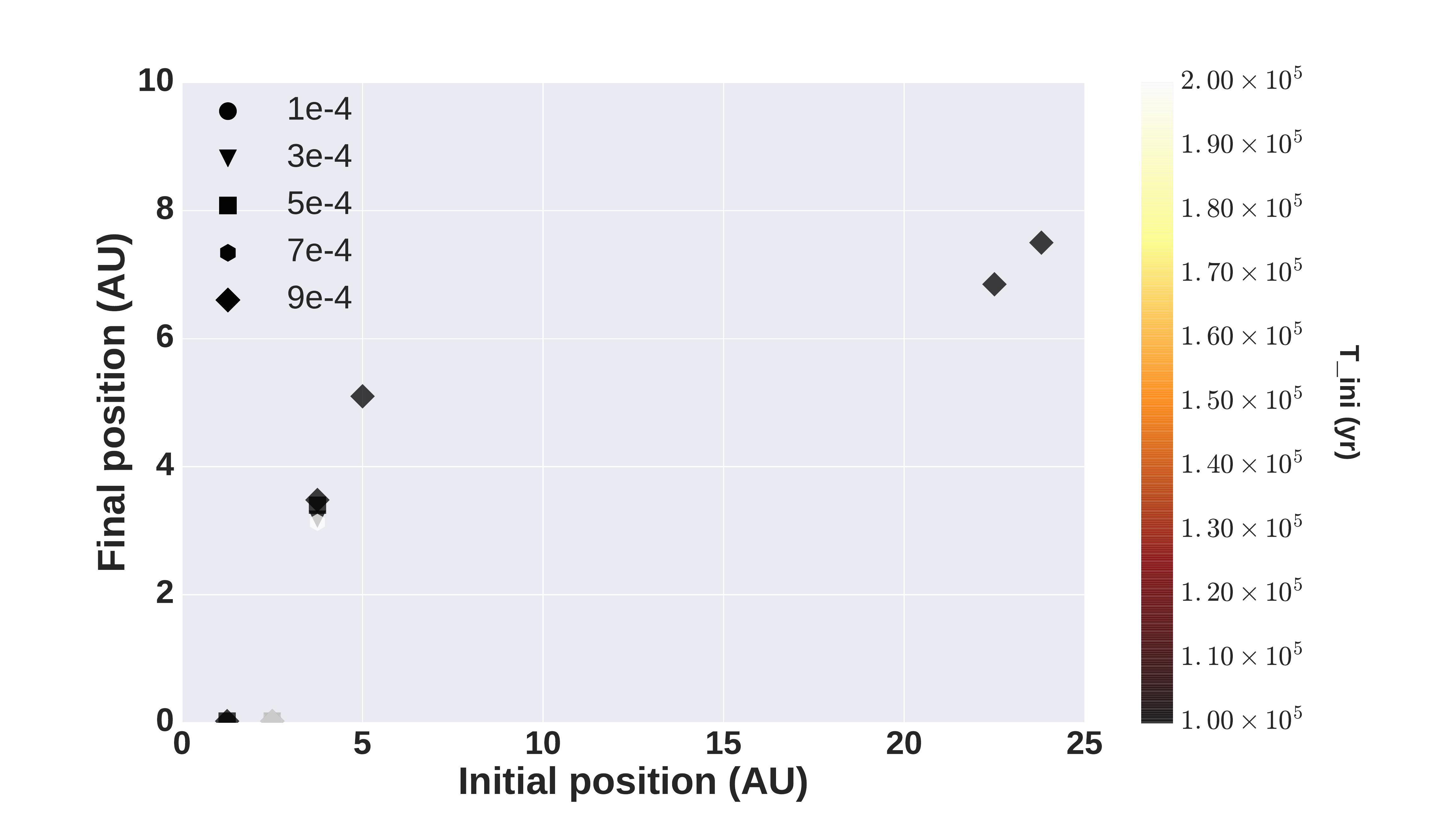}
   \caption{Same as Figure \ref{fig:sat1}, but with all constraints (chemical composition and core mass) applied.}
    \label{fig:sat2}
\end{figure}

\begin{figure*}
	\includegraphics[scale=0.2]{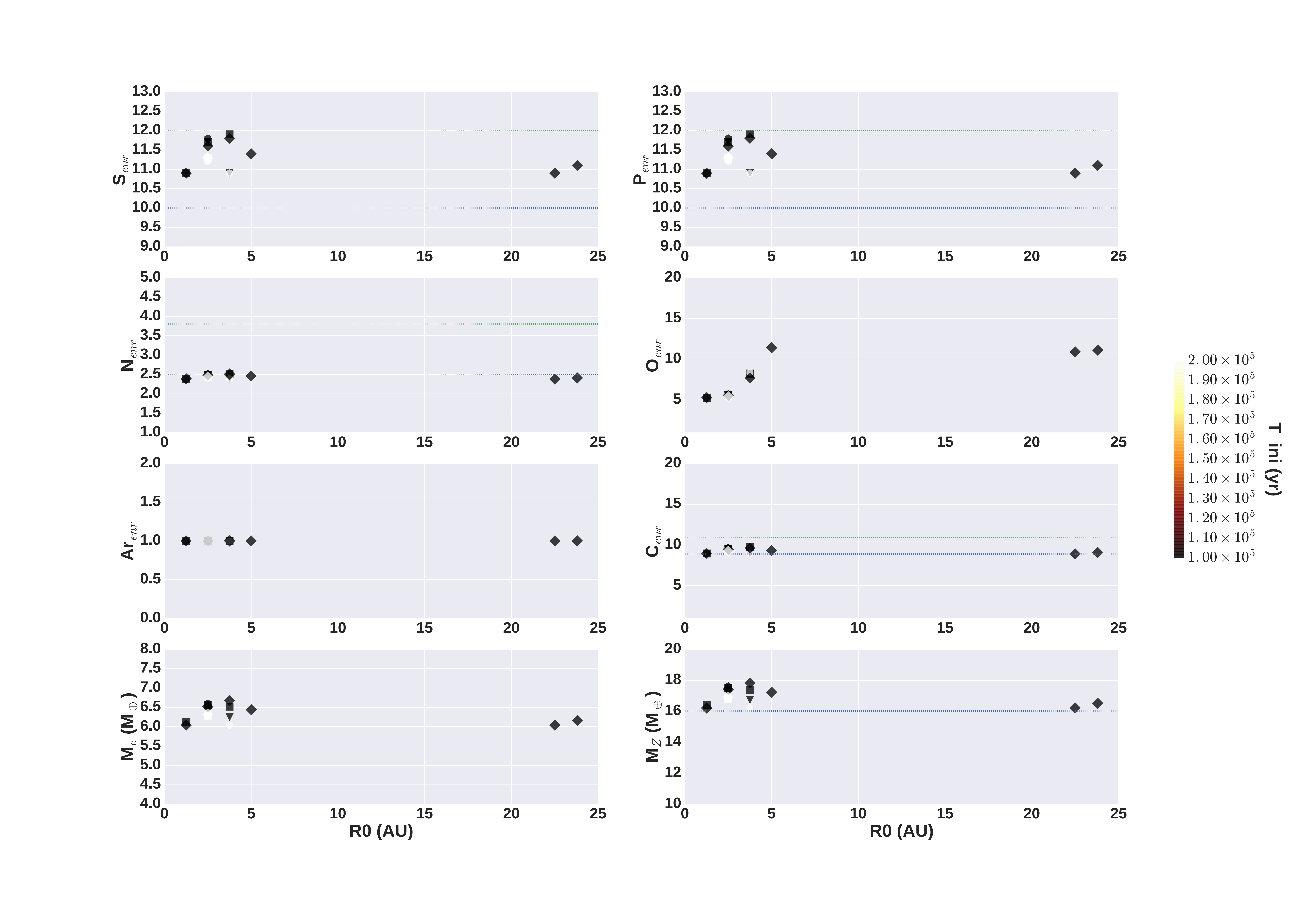}
   \caption{The enrichment factor of Sulfur, Phosphorus, Nitrogen, Oxygen, and Carbon predicted by this model for the case with no disk photoevaporation for Saturn. The dashed horizontal lines are the observational constraints. Bottom left panels show the predicted total amount of heavy elements in the planet. Core masses are now shown because all of these model have completely eroded cores.}
    \label{fig:sat3}
\end{figure*}

We follow the same approach as above for Saturn. Fig. \ref{fig:sat1} shows all of the 10$^4$ Saturn mass planets found by the model, with no constraints applied. The unique information we have on Saturn's total amount of heavy elements can be used along with its carbon abundance to {severely reduce the numbers of matches, keeping only 0.3\% of the initial population (using each of the constraints individually will keep around 1\% of the population).} The final fitting planets and their properties are shown in Figs \ref{fig:sat2} and \ref{fig:sat3}. {Again, including the Nitrogen abundance will leave almost no matches, so we choose to ignore it for the same rational highlighted in the Jupiter case.}  Interestingly, it seems that there are two possible {distinct} formation zones for Saturn : between 1 to 5 AU, and 20 to 25 AU. The region between 5 and 20 AU is excluded because it leads to a low total mass of heavy elements. {Additionally, all of the models who fit the observations have a core erosion factor of 60\%, meaning only a small residual core for Saturn. This value is comparable that of Jupiter, and implies that the two planets can have comparable core masses (even though significantly lower values are allowed for Jupiter). {Finally, all of the matching planets start forming very early in the disk, no later than 2$\times$ 10$^5$ yr, due to the constraints we have on $M_{Z}$ for Saturn.} The predictions from these models are shown in table \ref{t3}. {In Fig. \ref{fig:sat2} we also notice that no Saturn matching the constrains form exterior to Jupiter (Fig. \ref{fig:jup2}). This indicate that either our picture of the physical and chemical evolution of the disk needs to be revised, or that the no-photoevaporation model in not valid for our solar system. }}

\subsubsection{Single planet: Uranus/Neptune}
\begin{figure}
	\includegraphics[scale=0.2]{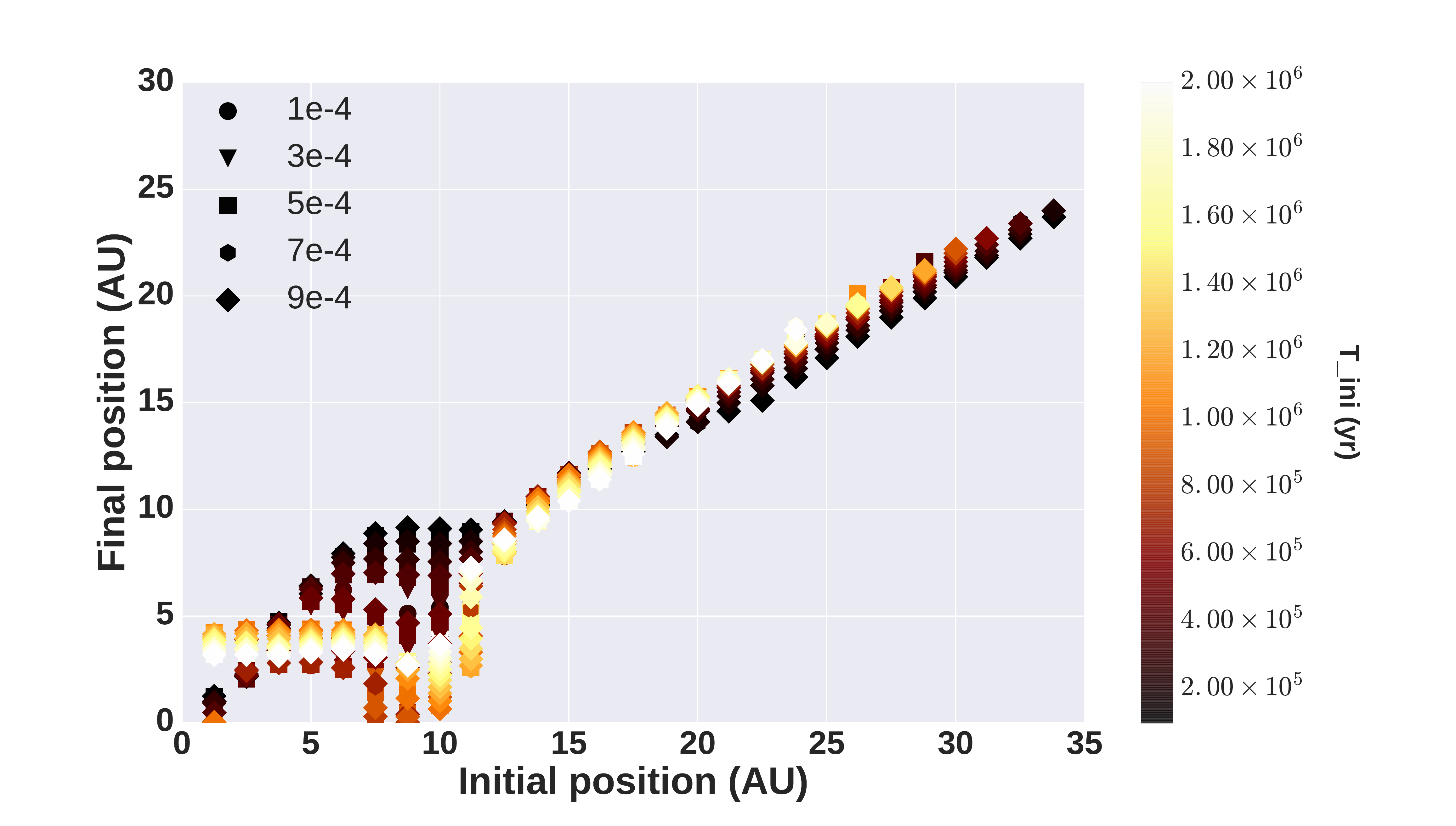}
   \caption{The final position (Rf) of Neptune mass planets as a function of the seed's initial position (R0) and injection time (T$_{ini}$). The geometrical forms are the seed's initial mass in M$_{\oplus}$. This is the case with no disk photoevaporation, and showing all planets with no constraints.}
    \label{fig:nep2}
\end{figure}

The same approach is also followed for a Neptune mass planet. The constraints we have for Neptune are its carbon abundance, D/H ratio, and informations on its interior structure.  Results for the case with no constraints applied is shown in Fig. \ref{fig:nep2}. This figure shows Neptune mass planets forming throughout the disk, which is not surprising since we are stopping the simulations when a Neptune mass is reached even if the disk is still young, so a large number of these Neptune are artifacts of the model, specially the ones in the inner disk. {This assumption however should have a minimal effect on the results since the planets founds (with and without applying constraints) are varied in term of their formation location and time, so if we were to relax this assumption and simply keep only the Neptunes who formed just before the disk's dissipation, we would still end up with the late forming subset of the population}. we first apply the chemical constraints (carbon and deuterium abundances), hence severally reducing the number of matches. Now only planets forming beyond 20 AU can be found (with a handful of planets around 3 AU that we choose to ignore for the above mentioned reason). These planets lay well on a straight line in the Initial-Final positions space. This is expected since Type I migration acting on these planets is linear. Accretion laws in this regime are also quasi linear to the position. Now we add constraints on the interior structure of these planets. \cite{helled} found, for cases where the interiors are dominated by water, total heavy elements masses of more than 12.8 M$_\oplus$ for Uranus and 15.2 M$_\oplus$ for Neptune. Including this constraint reduce the number of solutions all the way to zero, as long as the D/H ratio constraint was also present. {In other words, we can have matching planets if and only if either the D/H abundance or interior structure constraint is applied. Using both will give no matches at all. While we are not trying explicitly to match the water abundance of Uranus/Neptune because it is unknown to date, the D/H ratio and interior structure of the planets are used as a proxy to constrain the amount of accreted water ices.} This imply that this family of core accretion models are unable to conciliate the D/H ratio of Uranus \& Neptune with their interior structures. This paradox was noted initially by \cite{feucht}. \cite{ali-dibb} proposed as a solution the formation of Uranus and Neptune on the CO iceline from predominantly CO ices. This would also explain the seemingly very high ice/rock ratios in these planets interiors \citep{nettel}. {This possible solution is not fundamentally incompatible with the models presented here (and classical core accretion scenarios in general), and can be seen as an extension to the models. A model taking into account this mechanism self consistently in the framework of pebbles accretion should be interesting and is left for future work.}

\subsection{With photoevaporation}
Now we investigate the effect of our simplistic external photoevaporation on the chemical and physical evolution of the disk. We ran the simulations with the parameters specified in table \ref{t2} with only one planet in the disk, and we stop the simulations when either the planet reaches the mass of Jupiter/Saturn/Neptune or the disk photoevaporate completely. The FUV fluxes values were chosen in such a way to disperse the disk on {timescales in the order of 2-5 Myr.}
\subsubsection{Single planet: Jupiter}
The results of these simulations are presented in Fig. \ref{fig:jwipo1} and Fig. \ref{fig:jwipo2}. Figure \ref{fig:jwipo1} shows the distribution of Jupiter mass planets found by the simulations with no constraints applied other than mass, for different FUV rates. It is to be compared with Fig. \ref{fig:jup1}, showing the case with no photoevaporation at all. {We can first notice the expected general distribution trend similarities between the no photoevaporation and the 2 weakest FUV cases. In the two cases with significant PE, the early forming planets (with dark colors in the plot) are below the late forming planets (light colors) on the plot. This is due to an interplay between the disk structure, planets accretion rate and migration. Photoevaporation will decrease the planet gas accretion rate later in the disk, while increasing its metallicity, leading to the planet migrating less before it reaches Jovian mass. Planets forming before photoevaporation has taken over will thus migrate inward further. \\
Figure \ref{fig:jwipo2} shows the population of Jupiters that meet all of the constraining criteria {(Nitrogen abundance included, in contrast with the no-photoevaporation case)}. The most interesting cases are for FUV = $2 \times 10^{-9}$ and $3 \times 10^{-9}$, where this population's R0 extends sporadically all the way in to 10 AU. This is of course due to photoevaporation that increased the chemical enrichment of all {elements throughout the disk allowing planets with chemical compositions compatible with Jupiter to form much closer to the sun than in the case without PE.} Now we focus only on this subset of planets forming between 10 to 20 AU. These are found to be in two categories: those forming in a disk with FUV = $2 \times 10^{-9}$ all have a core erosion of 60-80\% and chemical enrichment factor due to PE of around 1.75, and the rest forming in a disk with FUV = $3 \times 10^{-9}$ have a core erosion of 60\% and chemical enrichment due to PE of around 1.80. Detailed informations and predictions of these simulations are shown in Fig. \ref{fig:jwipo3} including planets for all FUV values. We notice that prediction for the Oxygen abundance in Jupiter in the cases with PE is found to be very close to the value without PE, 3-4 $\times$ solar value. 
However, the two cases (with and without PE) seem to be, in principle, distinguishable by their core mass and total mass of heavy elements. The case with PE predict a possibly higher M$_Z$ value up to 23 $M_\oplus$, higher than the no-PE case. This is due to to the double origin of heavy elements in the planet: the core and the metals heavy (due to PE) envelope. The case with PE predict also a possibly higher M$_c$, where significantly more matches have M$_c$ in the order of 5 $M_\oplus$ than the case with no PE. This is simply because we need to erode less material from the core to reproduce the chemical composition of Jupiter if most of the enrichment is imported by the chemically evolved gas.\\
Looking at Fig. \ref{fig:jwipo3}, we can notice that a significantly larger number of matches (specially in the inner disk) can be obtained if either slightly less amounts of S and P were accreted, or more carbon. This implies that, if the PE model is correct and if Jupiter formed in the inner disk, slightly less than a 100\% of P and S were in refractory forms (to allow the accretion of less P and S solids), or carbon in refractories made up more than 80\% of the solar carbon abundance, which is unlikely from observations. A significant (in the order of 20\%) volatile fraction of P and S is however not excluded from observations or theory \citep{pasek2005}. This means that, even in the PE model, most of the Sulfur and Phosphorus in the protosolar nebula should have been in refractories.}

\subsubsection{Saturn}
We now run the same simulations as above but for Saturn. In this case, simulations are stopped when either the disk has fully photoevaporated or a Saturn mass planet has formed. We use the same FUV fluxes as for Jupiter. Figure \ref{fig:satwipo1} shows the population of Saturn mass planets with no constrains applied. It should be compared with Fig. \ref{fig:sat1} showing the case with no photoevaporation. The distributions found in both cases are similar, with the PE population more dispersed in the R0-Rf space, as seen for Jupiters and for the same reasons. Fig. \ref{fig:satwipo2} shows the population of Saturns with all constrains applied. Interestingly, this population is spread throughout the R0-Rf space, with no empty region between 10 and 18 AU as in the case without PE. This is again due to photoevaporation contributing to the total amount of heavy elements in the planet's envelope, stopping it from falling below the lower limit. This of course will lead to a lower core erosion factor, where these simulations show that in this case {it can be as low as 40\%, even though a significant population with 60\% core erosion still exist}. Predictions from these models are shown in Fig. \ref{fig:satwipo3}. {The planets forming below 5 AU are unfavored even though they are shown in the plots, because they might be incompatible with the bulk oxygen abundance of Saturn indirectly inferred from CO measurements \citep{mousis2014, wang2015}.} Interestingly, it seems that the best indicator to distinguish the with and without PE models {is the solid core and total amount of heavy elements in the planet. The PE case predict a residual solid core with mass between 7 and 15 $M_\oplus$, which should in principle be distinguishable from the 6-7 $M_\oplus$ in the no PE case, even though degeneracy exists. M$_Z$ predicted by the PE case is between 15 and 25 $M_\oplus$, again significantly larger than the 16-18 $M_\oplus$, predicted by the no PE case. This implies that, we can distinguish observationally between the two models for Saturn if and only if M$_c$, M$_Z$ larger than respectively 10, 20 $M_\oplus$ is detected. It is also remarkable that most of the cases with the 40\% core erosion factor (corresponding the upper population in the M$_c$ and M$_Z$) form earlier than the 60\% core erosion population. This implies that, if indeed a high M$_c$ and M$_Z$ values were detected in Saturn (in favor of the PE case), Saturn much have started forming very early in the protosolar nebula. {Finally, in contrast with the case without photoevaporation, here we do find Jupiter-Saturn couples in the right order and compatible with both planets full sets of constraints. The possible formation zone of both planets extends all the way from the inner disk to 35 AU, allowing Saturn to form outside of Jupiter.}}

\subsubsection{Uranus \& Neptune}
Next we run simulations that stop when reaching Neptune mass planets or the disk is photoevaporated. The case with no constrained applied is shown in Fig. \ref{fig:nwipo1}. The Neptunian planets population found is very similar to the case with no photoevaporation. The population is also almost identical for different FUV rates. This is expected since photoevaporation should have a much lesser effect on the low mass Neptunian planets who's formation is much quicker and necessitate less gas than Saturn or Jupiter mass planets. \\
When applying all of the constraints (Carbon abundance, D/H ratio and interior structure), no matches were found at all. This is due, as in for the no PE case, to classical formation models being unable to account at the same time for Uranus \& Neptune interior structure and their chemical composition \citep{ali-dibb}.

\subsubsection{Multiple planets: Jupiter and Saturn}
In this section we investigate the case of 2 planets in the disk. The goal is to check which set of initial conditions lead to 2 giant planets with Jupiter and Saturn masses that cross the 2:3 MMR, and if their chemical composition is compatible with Jupiter and Saturn. {The reason for including this criteria is that a large of the Jupiter-Saturn matching pairs we find do end up significantly inward of their current orbits. It is then important to check if these planets cross this MMR (thus reversing their migration) before the dispersal of the disk.} In each simulation, a pair of seeds are injected into the disk and then left to evolve while checking at each timestep if they crossed any of the resonances. In this case, their migration would be turned off, but their masses can still evolve. In table \ref{restab} we show planets pairs that meet the dynamical criteria. More matching pairs of planets in the case with PE. It seems that PE allow planets forming later in the disk (1.3-1.5 Myr) to cross their mutual 3:2 resonance. This is because migration rates are linear to the disk's surface density affected by PE. However, None of these pairs of planets meet the full set of chemical constraints. {It is possible that this result is hinting toward the importance of the fast type III runaway migration \citep{masset2004,walsh2011} not included in our model, or more realistic disks \& type I and II migrations modeling \citep{kley2012,baillie2015}.}

\renewcommand\arraystretch{1.2}
\begin{table}
\begin{center}
\caption{The predictions for the Nitrogen \& Oxygen abundances, core masses and envelopes metallicity made from these models for Jupiter and Saturn.}
\small
{\begin{tabular}{lccccc}
\hline
\noalign{\smallskip}
				& \multicolumn{2}{c}{Jupiter}				& \multicolumn{2}{c}{Saturn}	 	\\
Observable			& No PE			& PE		& No PE			&  With PE 	\\
\hline
N				&  - 		& 2-4					& -		& 2-3						\\
O	& 3-4		& 3-4					& 5-12			& 	7-13	 				 \\		
M$_c$ (M$_\oplus$)				& 2 or 5		& 1.5 or 4-5					& 6-7		& 7-15						\\		
M$_Z$ (M$_\oplus$)				& 15-17  	& 15-24					& 16-18			& 15-27						\\		
\hline		
\end{tabular}}\\
\label{t3}
\end{center}
\end{table}

\renewcommand\arraystretch{1.2}
\begin{table}
\begin{center}
\caption{The relationship between the different observational constraints and the model's free parameters.}
\small
{\begin{tabular}{lccccc}
\hline
\noalign{\smallskip}
Constraint			& Free parameter \\		
\hline
M$_c$ - M$_Z$				&  R$_0$/t$_0$ $^{\rm(a)}$ 	\ \& \ 	$\dot{M}_{FUV}$	$^{\rm(b)}$ 					\\
S - P abundances  	& M$_c$ - M$_Z$		$^{\rm(c)}$								 \\		
N - Ar	abundances			& R$_0$/t$_0$ $^{\rm(d)}$											\\		
C	abundance			& C$_{ref}$		$^{\rm(e)}$							\\		
D/H			& M$_c$ - M$_Z$ \&  D/H$_{ice}$  $^{\rm(f)}$ 						\\		

\hline		
\end{tabular}}\\
$^{\rm(a)}$ {Through the isolation mass, controlled by the disk's location and time dependent aspect ratio.}\\	
$^{\rm(b)}$ A Photochemically enriched disk will lead to higher M$_Z$ and less core erosion.\\	
$^{\rm(c)}$	Assuming S and P are present in disk as purely refractories.			\\
$^{\rm(d)}$ Through the time dependent icelines location.	\\
$^{\rm(e)}$ If the more volatile Nitrogen abundance is not used as constraint.\\
$^{\rm(f)}$ Since the D/H in Uranus \& Neptune depend on the amount of accreted \& well mixed water ices and their D/H ratio.

\label{summary}
\end{center}
\end{table}

\renewcommand\arraystretch{1.2}
\begin{table}
\begin{center}

\caption{The pairs of Jupiter-Saturn planets who are compatible with their chemical compositions and cross their mutual 2:3 MMR during their evolution. All of these are in the case with PE.}
\tiny
{\begin{tabular}{lcccccc}
\hline
\noalign{\smallskip}
T$_{ini}$ Jup (yr)			& T$_{ini}$ Sat 	& R$_{ini}$ Jup (AU)		& R$_{ini}$ Sat		&  R$_{fin}$ Jup	 &  R$_{fin}$ Sat 	\\
\hline
& \multicolumn{4}{c}{FUV = $2\times 10^{-9}$} \\
\hline
9.0$\times 10^5$				& 1.1$\times 10^6$		& 5					& 10	 & 0.125	& 0.175		\\		
1.5$\times 10^6$				& 2.1$\times 10^6$		& 2.5					& 7.5	 & 0.1	& 0.125		\\		
1.5$\times 10^6$				& 2.1$\times 10^6$		& 5					& 7.5	 & 0.1	& 0.125		\\		
\hline	
& \multicolumn{4}{c}{FUV = $3\times 10^{-9}$} \\
\hline
9.0$\times 10^5$				& 1.1$\times 10^6$		& 2.5					& 10	 & 0.125	& 0.175		\\		
9.0$\times 10^5$				& 1.1$\times 10^6$		& 5					& 10	 & 0.125	& 0.175		\\		
1.3$\times 10^6$				& 1.9$\times 10^6$		& 2.5					& 7.5	 & 0.1	& 0.125		\\		
1.3$\times 10^6$				& 1.9$\times 10^6$		& 5					& 7.5	 & 0.1	& 0.125		\\		
1.5$\times 10^6$				& 2.1$\times 10^6$		& 2.5					& 7.5	 & 0.1	& 0.125		\\		
1.5$\times 10^6$				& 2.1$\times 10^6$		& 5					& 7.5	 & 0.1	& 0.125		\\		
\hline

\end{tabular}}\\

\label{restab}
\end{center}
\end{table}

\section{Discussion \& Conclusions}

\subsection{Caveats \& perspectives}
Our model incorporate a large number of different disk and planets physical and chemical effects. It is inherently simplified for numerical complexity reasons. The following caveats can hence be identified:
\begin{itemize}
\item The major caveat of this work is the simplicity of the disk external photoevaporation model. For more robust and quantitative results, proper dynamical coupling of a viscously evolving disk to photoevaporation and planets formation is needed and should be addressed in future works. The case of internally photoevaporated disk is also relevant and should be explored.
\item Another possible caveat is the unevolving disk chemistry we assumed. Even though our disk chemistry is based on small bodies and disks observations, it is unlikely that this chemistry was static and unchanged during the disk lifetime. However to date no chemical model was able to fully reproduce the chemical abundances of today's small bodies from first principles. Nonetheless, a more sophisticated disk chemistry can bring valuable insights on the formation of giant planets.
\item In interpreting the chemical composition of the giant planets atmospheres as the bulk elemental abundance, we are implicitly assuming these planets to be fully convective. Even though this seems as the most likely case, it is still the subject of an ongoing debate \citep{guillot2014}. 
\item Many of the disk parameters (dust and pebbles metallicities) and planets parameters (envelope density and opacity) were fixed to a constant value throughout this work to keep the problem tractable. Exploring these parameters should shed more lights on the formation of giant planets, specially exoplanets, and is left for future work. 
\item In this model we implemented a very simple yes/no indicator to check if Jupiter/Saturn crossed their mutual 2:3 mean motion resonance. For future works however it is important to track the full dynamical evolution of these planets, {along with type III migration,} and to include Uranus and Neptune. 
\item {This work used the chemical composition of three comets (1P/Halley, 67P/C-G and 81P/Wild) as indicators for the disk's chemistry. In reality, dynamical mixing of the cometary reservoirs have probably meant that both OCCs and JFCs probably sample material formed at a wide variety of orbits \citep{altwegg,brasser}. A deeper understanding of the statistics of comets chemistry is key to improve upon this work.}
\item {The models presented in this work assume that the heavy elements in the planet's envelope originates as either material eroded from the core or delivered from a chemically evolved gas. Heavy elements however can also be delivered into the envelope from late contamination by planetesimals. This phenomena however is inefficient. \cite{guillot2000,matter} for example found that the late heavy bombardment could have increased the volatiles enrichment in the atmospheres of Jupiter and Saturn by factors of respectively 0.033 and 0.074. }
\item {In this model, planet and disk evolution are calculated separately. We do not take into account the effect of solids removed from the disk into the forming planet on the structure of the disk. This however should should be a second order importance to the effect of the planets itself on the disk. The reason is the nature of pebbles accretion, were pebbles form continuously in the outer disk and drift inward till they get accreted by the core. The structure of the pebbles disk is therefore not being affected by accretion.}
\item {Another assumption concerning the disk structure is the constant accretion rate throughout the disk. This is unlikely to be true in the outer parts of the disk where the viscous timescale is comparable to the disk lifetime. Quantifying this effect however necessitate modeling the viscous evolution of the disk, which not taken into account in the disk model of \cite{bitsch2}. A non uniform accretion rate might lead to disk structure affecting the evolution of planets (specially migration). Modeling these effects is left for future works.}
\end{itemize}

\subsection{Summary}
In this work we explored the chemical composition of giant planets atmospheres using a formation model that includes simplified pebbles \& gas accretion, type I and II migration, solar system chemistry observations, \& external photoevaporation, and planet core erosion. We ran population synthesis simulations for individual planets in disks with and without photoevaporation to understand the origins of the solar system giant planets chemical composition, in addition to simulations with both Jupiter and Saturn to check the compatibility of their chemical composition with their dynamical histories. Our conclusions can be summarized as follow:

\begin{itemize}
\item Since Sulfur and, more probably, Phosphorus were both present as mostly refractory phases throughout the protosolar nebula, and because the uncertainties on their abundances in Jupiter are low, they can hence be used to constrain the fraction of core mass that got eroded into its envelope. The other more volatile (hence location dependent) elements such as Argon and Carbon can be used to constrain where/when in the disk did Jupiter form. {The case with photoevaporation however hints toward a fraction of S and P in volatile phase, since this is the only possibility to form Jupiter in the inner 10 AU.} 

\item {The highest fraction of Nitrogen in refractories measured so far in today's solar system (14\% solar abundance) along with a 20 K N$_2$ condensation temperature are too low to explain the measured nitrogen abundance in both Jupiter and Saturn if no significant photoevaporation induced disk chemical enrichment took place. This implies that either the Nitrogen chemically evolved significantly in the disk or that it condensed at higher temperatures through trapping in clathrate-hydrate ices.}

\item In the case with no disk photoevaporation, Jupiter's chemical composition is only obtained if it formed beyond 24 AU. This model predicts an oxygen enrichment in the order of 3 times solar, a residual core of 2 or 5 M$_{\oplus}$ where 60-80\% of the original core was eroded, and a total mass of heavy elements in the planet (M$_Z$) in the order of 15-17 M$_{\oplus}$. Saturn on the other hand can form in a larger swath of the disk (1-5 and 20-25 AU). The model predicts a 40-60\% eroded core for this planet with M$_c$ around 6-7 M$_{\oplus}$ and M$_Z$ in the order of 16-18 M$_{\oplus}$ and an oxygen enrichment between 5 and 12 times solar. {No single disk was found to form both planets in the right order with all their constraints applied, indicating that either different physics or chemistry are at play, or that this case is not a valid for our solar system formation.} In the case of Uranus \& Neptune, no set of parameters were found to match at the same time their chemical composition and interior structure constraints. 

\item {In the case with disk photoevaporation, a planet with Jupiter's chemical composition can form all the way in to around 8 AU, due to the disk's photoevaporation enriching it in heavy elements. The gas accreted by the planet is usually enriched due to PE by factors slightly less than 2 with respect to solar value. The rest of the metals in the envelope have to be provided by core erosion. These Jupiters have a residual core mass in the order of 1 to 3 M$_{\oplus}$, M$_Z$ in the order of 14-15 M$_{\oplus}$. This model predicts the same oxygen abundance for Jupiter as in the case with no disk PE. In the case of Saturn, planets respecting all the constraints can be created continuously throughout the disk. These are separated into two populations with core erosion factors between 40\% and 60\%, giving respective lower limits on M$_c$ and M$_Z$ of 7 and 15 M$_{\oplus}$, and upper limits of 15 and 25 M$_{\oplus}$. The predicted water enrichment for Saturn is in the order of 12-15 times solar abundance, even though values as low as 6 are not excluded. The cases with and without PE might then be distinguishable by M$_c$ and M$_Z$ of the giant planets, but degeneracies can be present. In contrast to the case without PE, pairs of Jupiter-Saturn respecting all of the constrains and forming in the right order were found in the case with PE. As in the case with no disk PE, no matches for Uranus \& Neptune have been found. This might be implying {a more complex formation mechanism for these planets similar to the CO ices enhancement proposed by \citep{ali-dibb}}.}

\item When checking for the pairs of Jupiter \& Saturn masses planets who crossed their mutual 2:3 resonance while ending up with compatible chemical compositions, non were found in the two cases with and without PE. This might be hinting to a major role of type III migration in our solar system history. 

\end{itemize}

\begin{figure}
	\includegraphics[scale=0.2]{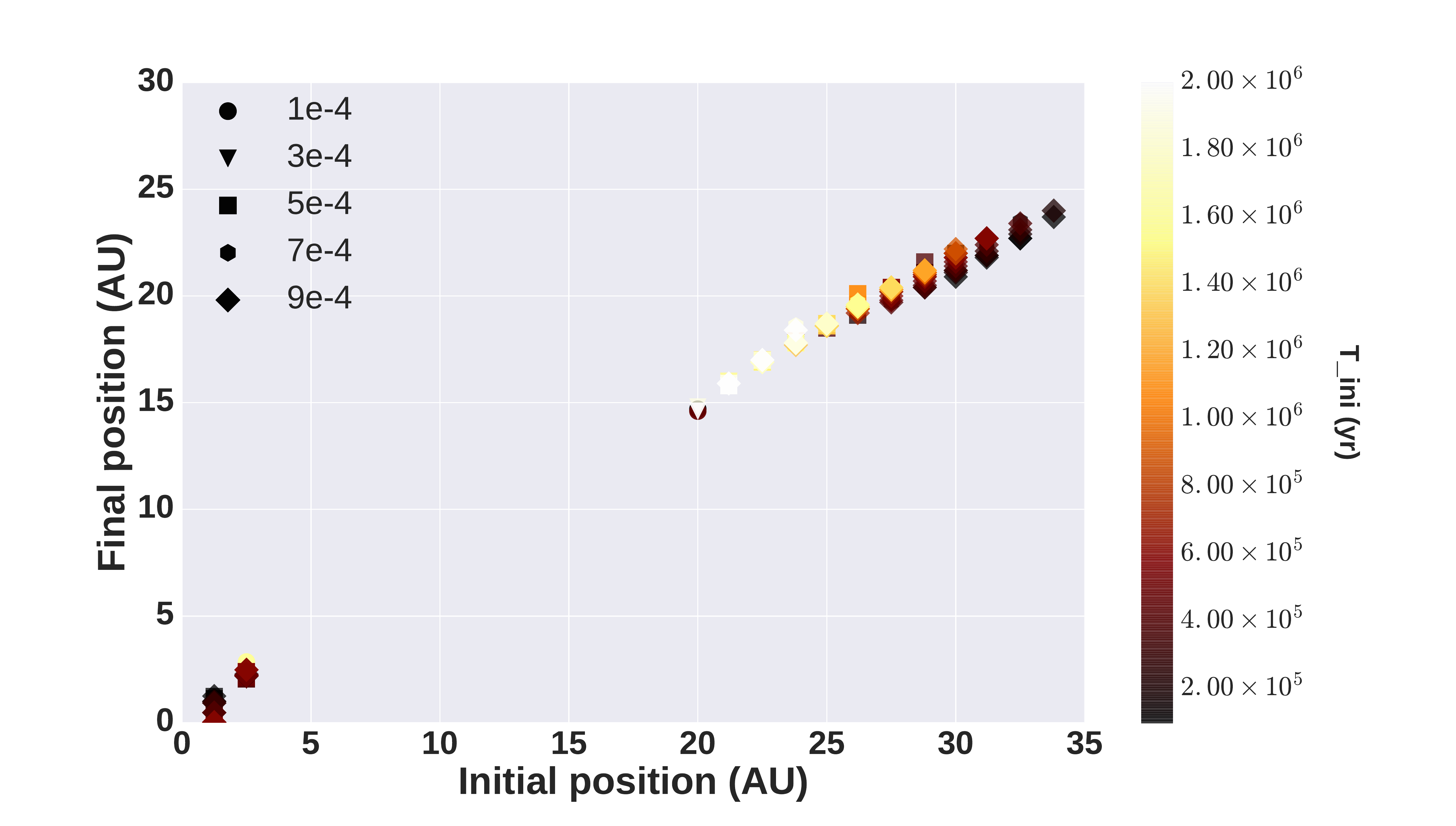}
   \caption{Same as Figure \ref{fig:nep2}, but with the carbon abundance and D/H constraints applied. Adding the informations on the interior structure will remove all matching planets.}
    \label{fig:nep1}
\end{figure}

\begin{figure*}
	\includegraphics[scale=0.2]{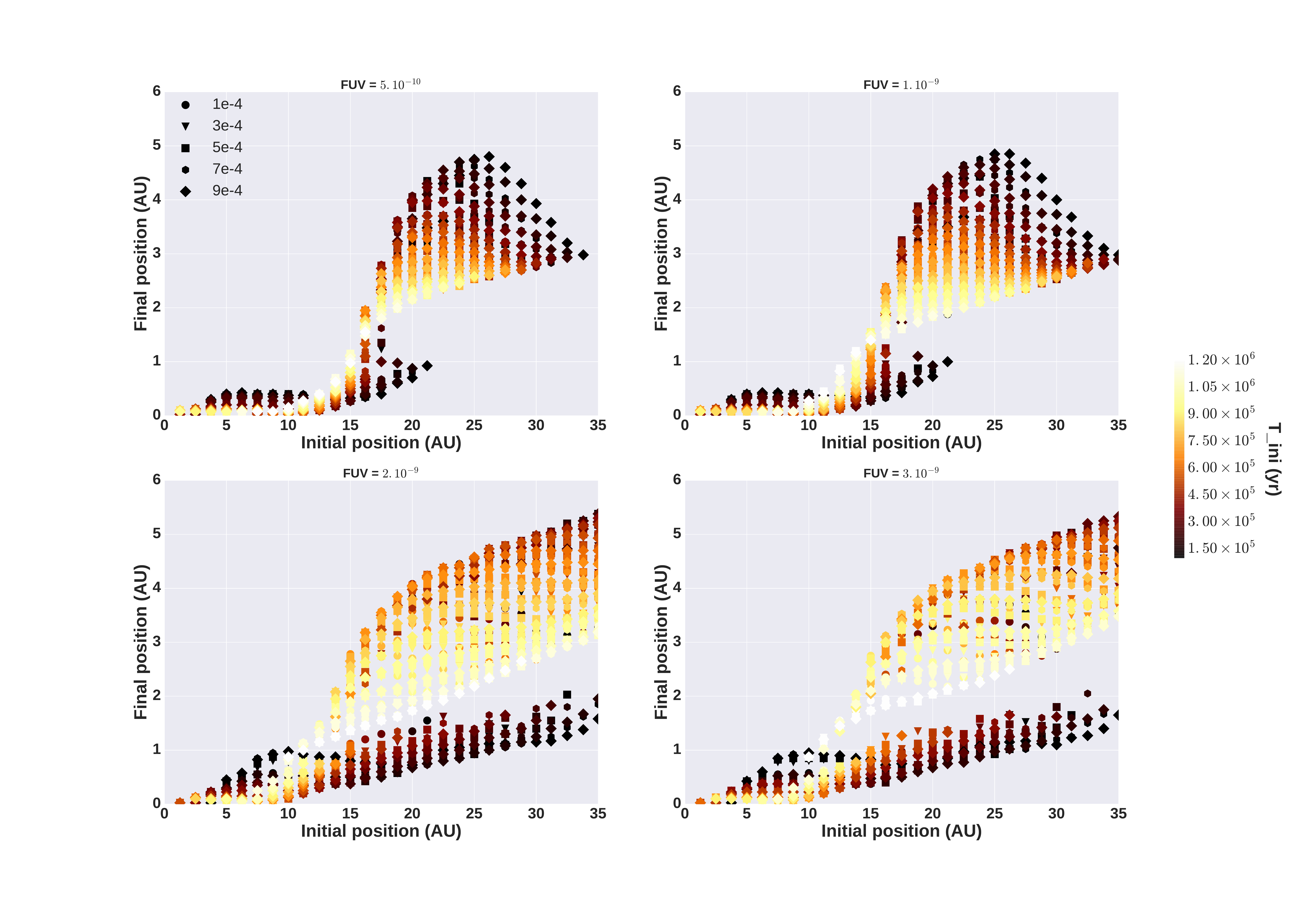}
   \caption{The final position (Rf) of Jupiter mass planets as a function of the seed's initial position (R0) and injection time (T$_{ini}$) and showing all planets with no constraints. The geometrical forms are the seed's initial mass in M$_{\oplus}$. This is for case with disk photoevaporation, where FUV is the value of $\dot{M}_{FUV}$ used.}
    \label{fig:jwipo1}
\end{figure*}

\begin{figure*}
	\includegraphics[scale=0.2]{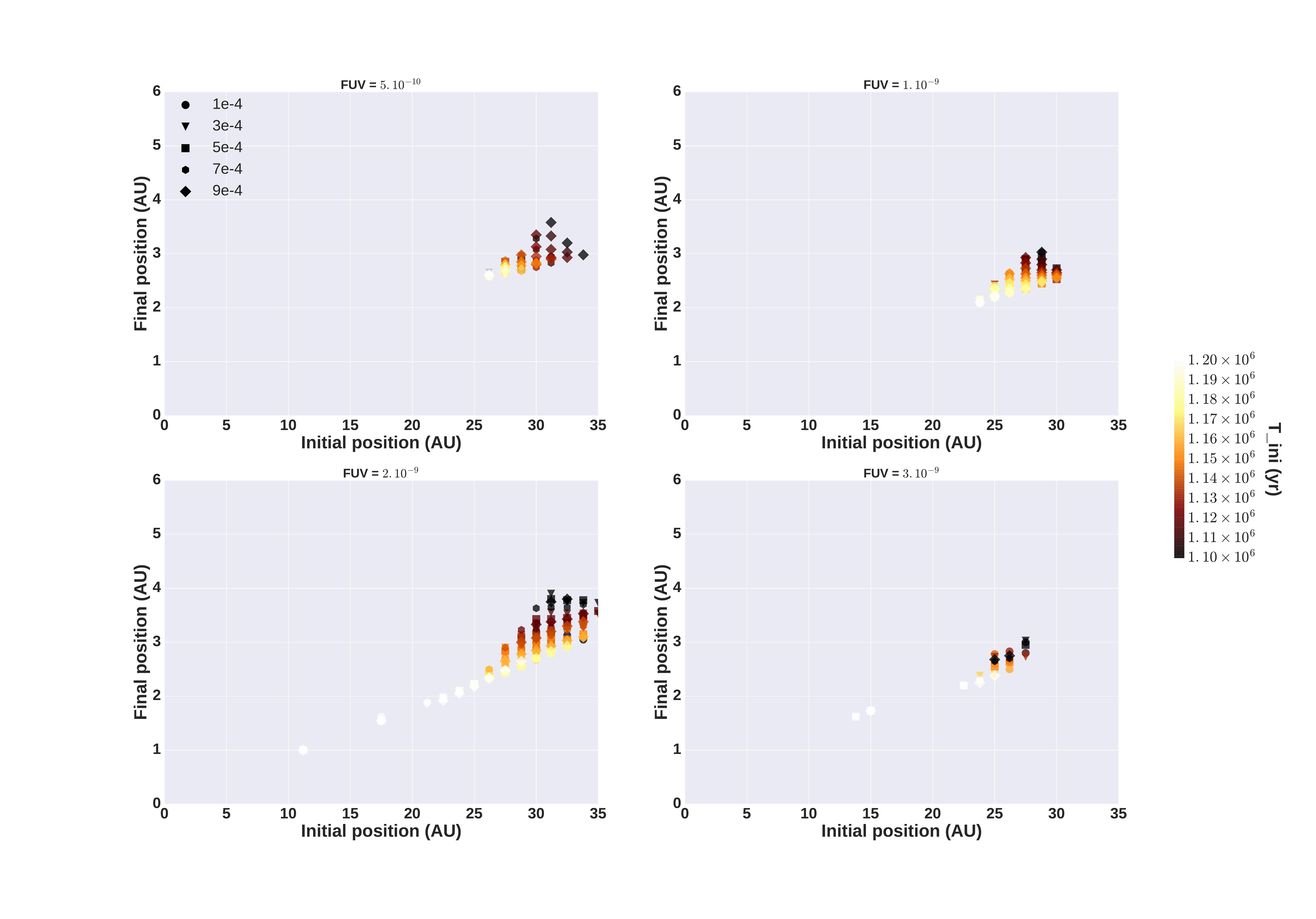}
   \caption{Same as Figure \ref{fig:jwipo1}, but with all constraints (chemical composition and core mass) applied. This is in the case with disk photoevaporation included.}
    \label{fig:jwipo2}
\end{figure*}

\begin{figure*}
	\includegraphics[scale=0.2]{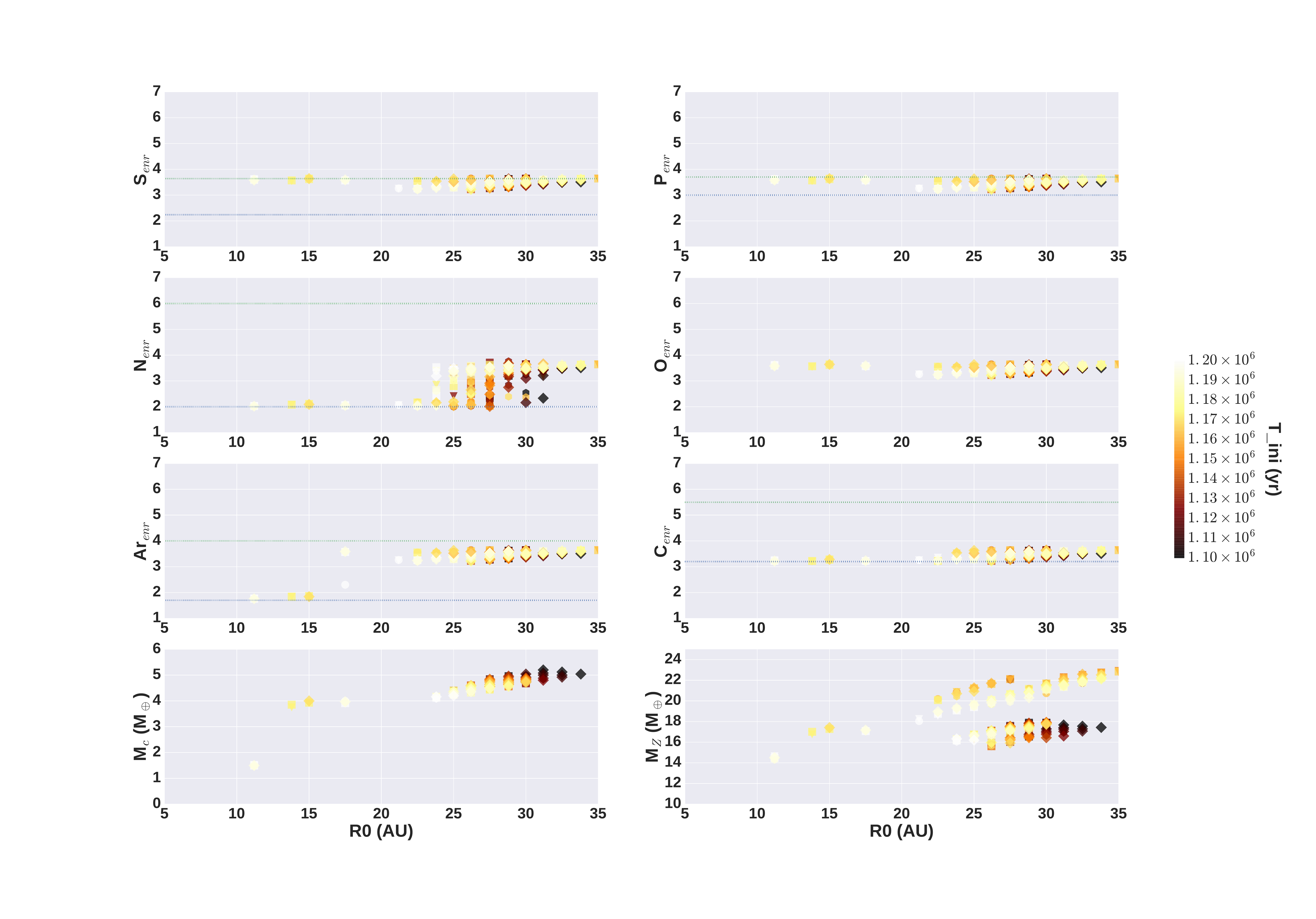}
   \caption{The enrichment factor of Sulfur, Phosphorus, Nitrogen, Oxygen, Argon and Carbon predicted by this model for the case with disk photoevaporation included for Jupiter. Bottom panels show the predicted core mass and total amount of heavy elements in the planet. The dashed horizontal lines are the observational constraints.}
    \label{fig:jwipo3}
\end{figure*}

\begin{figure*}
	\includegraphics[scale=0.2]{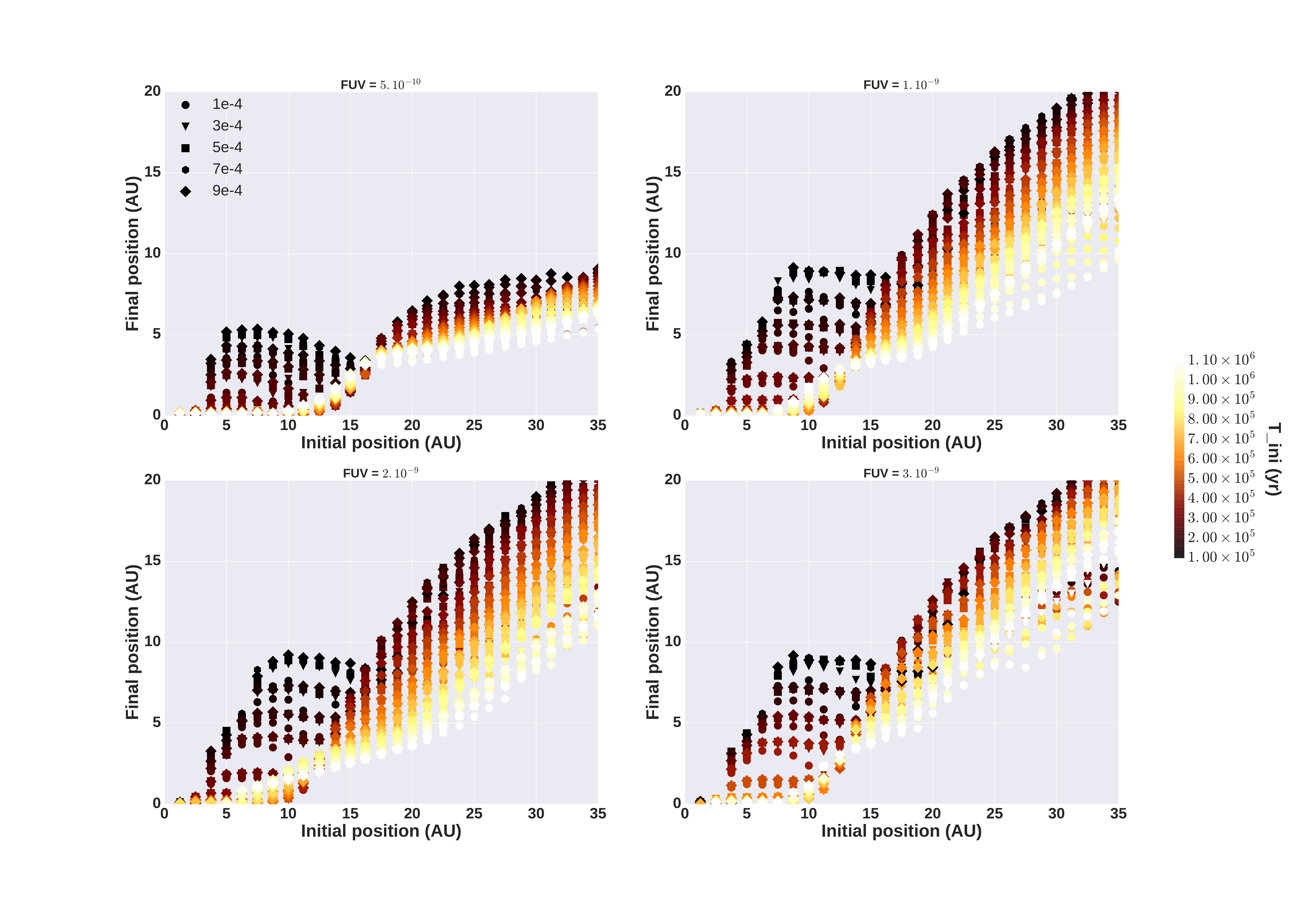}
   \caption{The final position (Rf) of Saturn mass planets as a function of the seed's initial position (R0) and injection time (T$_{ini}$) and showing all planets with no constraints. The geometrical forms are the seed's initial mass in M$_{\oplus}$. This is for case with disk photoevaporation, where FUV is the value of $\dot{M}_{FUV}$ used.}
    \label{fig:satwipo1}
\end{figure*}

\begin{figure*}
	\includegraphics[scale=0.2]{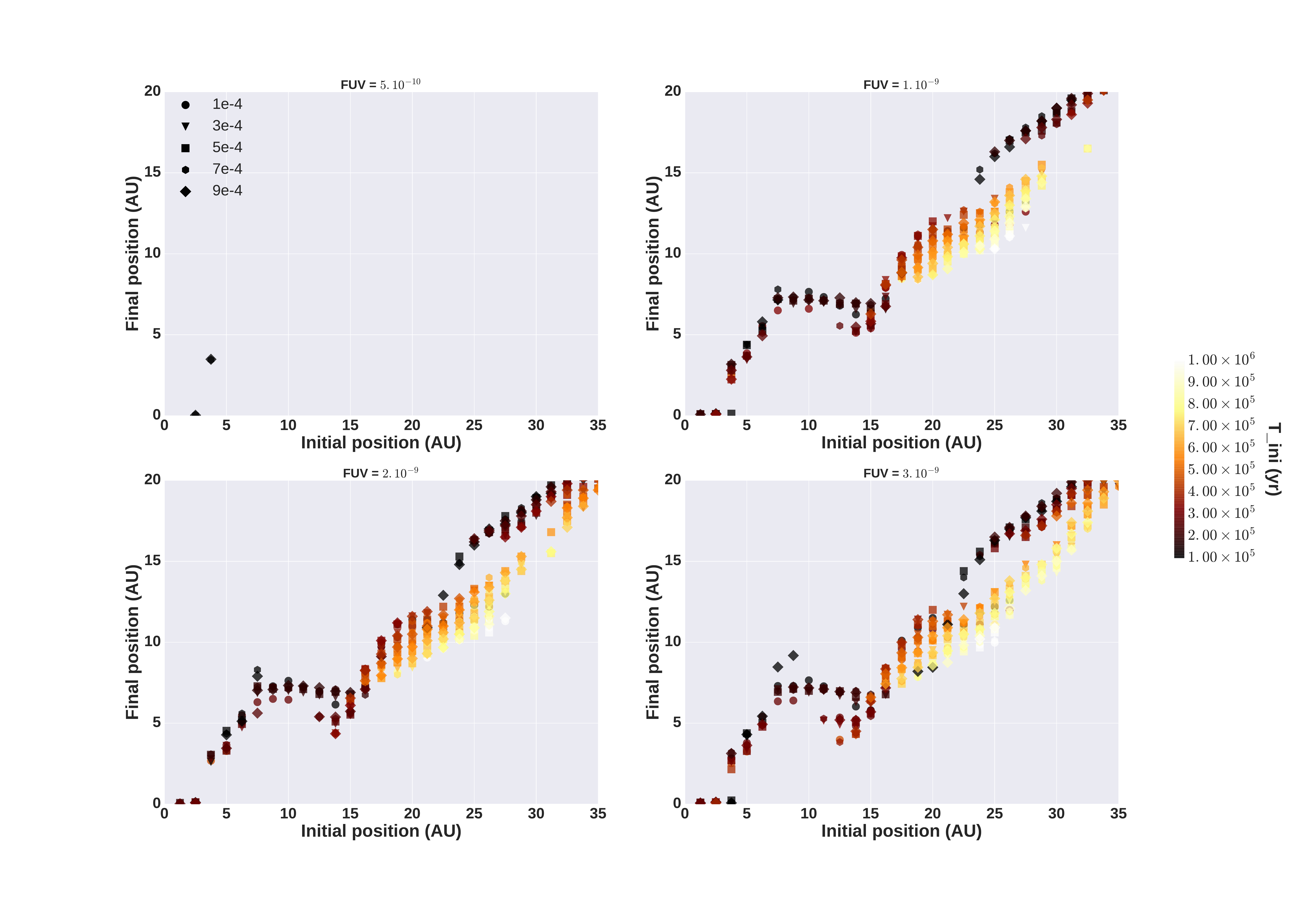}
   \caption{Same as Figure \ref{fig:satwipo1}, but with all constraints (chemical composition and core mass) applied. This is in the case with disk photoevaporation included. The two distinct populations seen correspond to the 40\% and 60\% core erosion factors. These values should hence be seen as lower and upper limits for the parameter being constrained.}
    \label{fig:satwipo2}
\end{figure*}

\begin{figure*}
	\includegraphics[scale=0.2]{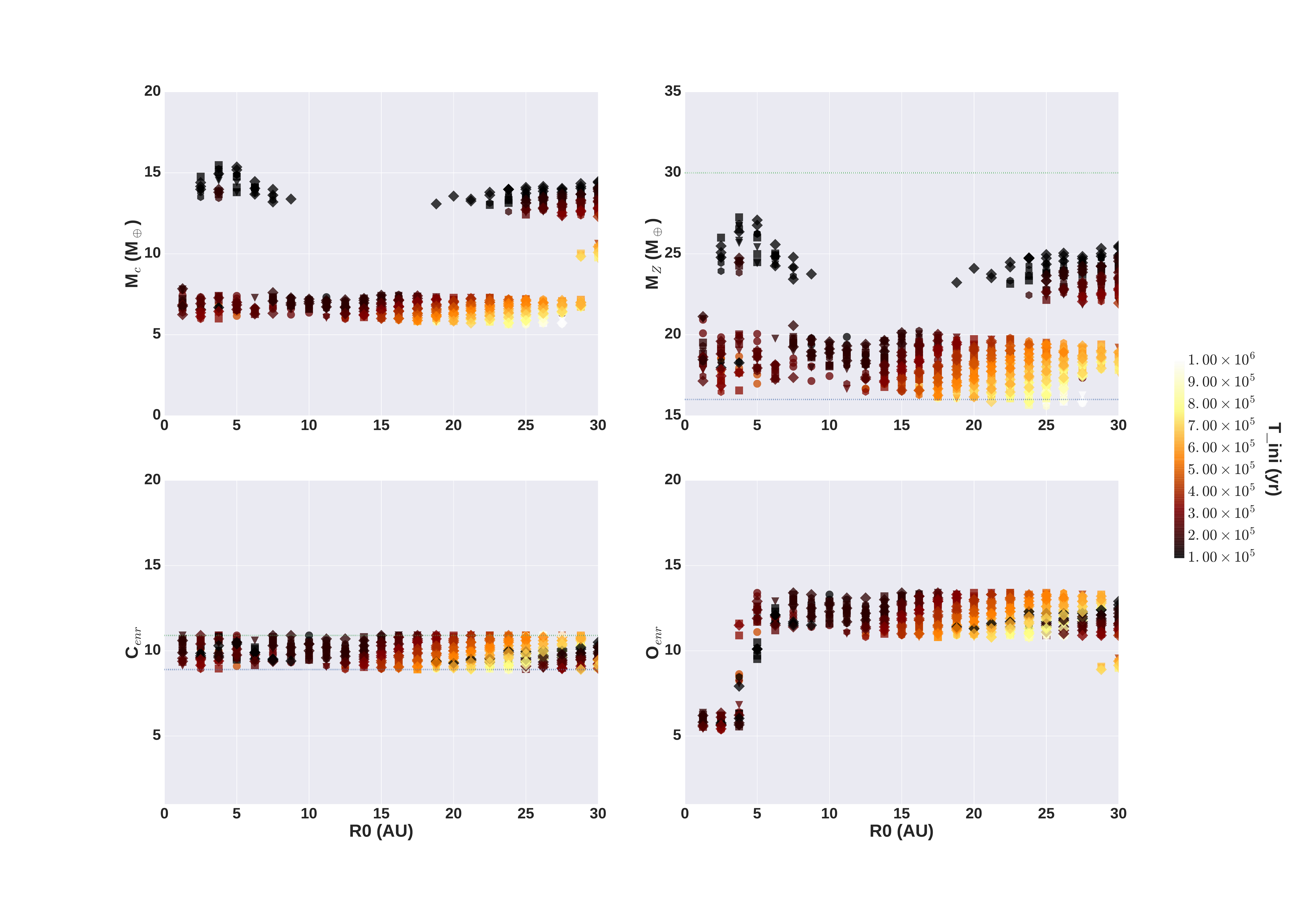}
   \caption{The enrichment factor of Oxygen, and Carbon predicted by this model for the case with disk photoevaporation for Saturn. Top panels show the predicted total amount of heavy elements in the planet and their core masses. The dashed horizontal lines are the observational constraints. }
    \label{fig:satwipo3}
\end{figure*}
\begin{figure*}
	\includegraphics[scale=0.2]{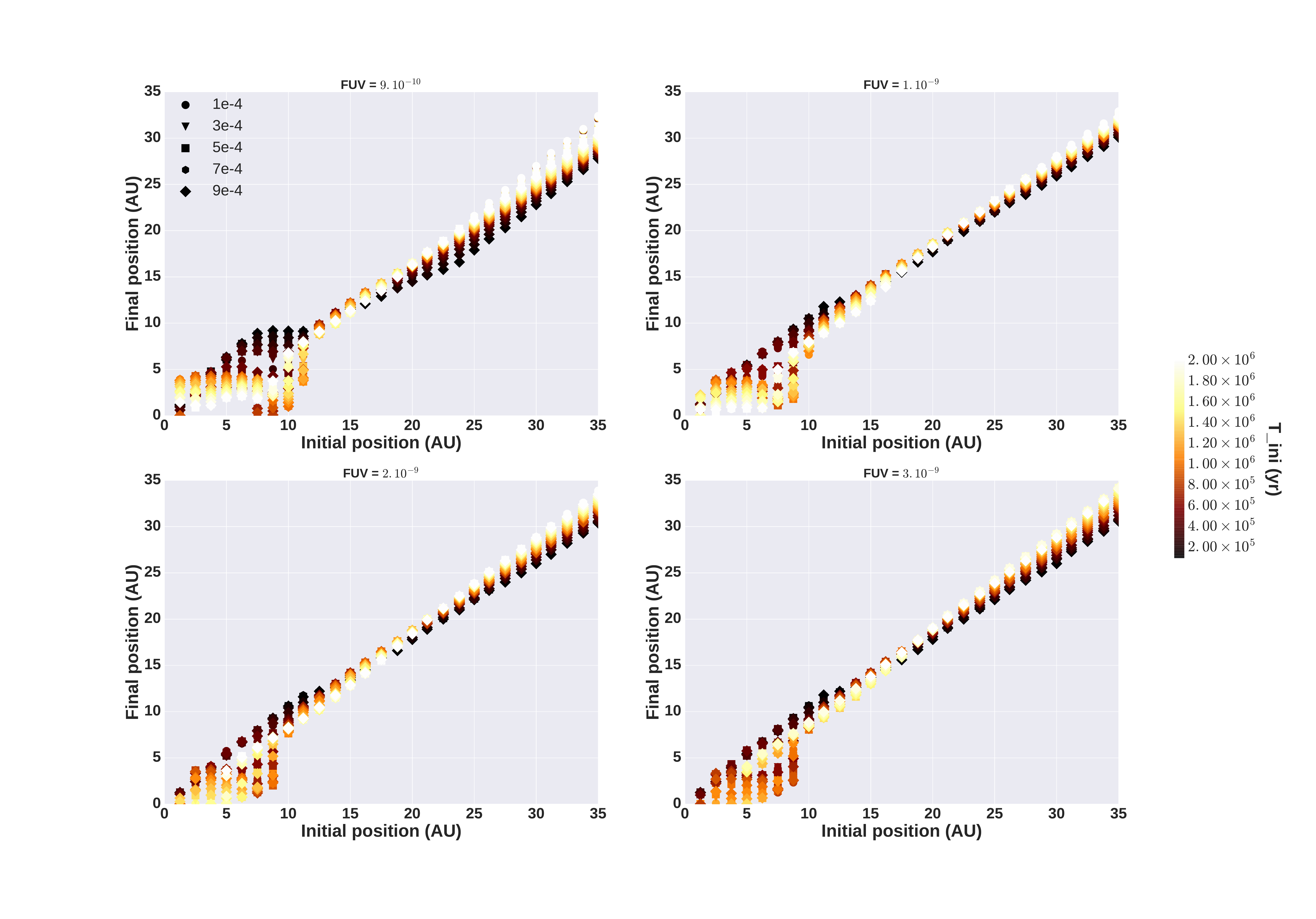}
   \caption{The final position (Rf) of Neptune mass planets as a function of the seed's initial position (R0) and injection time (T$_{ini}$) and showing all planets with no constraints. The geometrical forms are the seed's initial mass in M$_{\oplus}$. This is for case with disk photoevaporation, where FUV is the value of $\dot{M}_{FUV}$ used.}
    \label{fig:nwipo1}
\end{figure*}



\section*{Acknowledgements}

Data was analyzed using Python's Pandas package. We thank B. Bitsch for providing us with the disk model fits, and answering our many questions. We thank the anonymous
referee for useful comments that significantly improved the manuscript. Special thanks go to the Centre for Planetary Sciences group at the University of Toronto for useful discussions.








\appendix


\bsp	
\label{lastpage}
\end{document}